\documentclass[aps, prl, reprint, amsmath, amssymb, superscriptaddress, floatfix]{revtex4-1}

\usepackage{bm}
\usepackage{bbm}
\usepackage[retainorgcmds]{IEEEtrantools}
\usepackage{graphicx}
\usepackage{mathrsfs}
\usepackage{braket}
\usepackage{amsmath}
\usepackage{amsfonts}
\usepackage{amssymb}
\usepackage{color,xcolor}

\usepackage{times}

\usepackage{nicefrac}
\usepackage{ragged2e}
\usepackage{float}
\usepackage{tikz}
\usepackage{bbold}

\usepackage[colorlinks=true,linkcolor=blue,urlcolor=blue,citecolor=blue,pdfusetitle]{hyperref}

\usepackage{blindtext}

\newtheorem{result}{Result}
\usepackage{dsfont}

\usepackage[normalem]{ulem}

\begin{document}

\title{Deterministic Equations for Feedback Control of Open Quantum Systems}
\date{\today}
\author{Alberto J. B. Rosal}
\email{abezerra@ur.rochester.edu}
\affiliation{Department of Physics and Astronomy, University of Rochester, Rochester, New York 14627, USA}
\affiliation{University of Rochester Center for Coherence and Quantum Science, Rochester, New York 14627, USA}
\author{Patrick P. Potts}
\affiliation{Department of Physics and Swiss Nanoscience Institute,
University of Basel, Klingelbergstrasse 82 CH-4056, Switzerland}
\author{Gabriel T. Landi}
\affiliation{Department of Physics and Astronomy, University of Rochester, Rochester, New York 14627, USA}
\affiliation{University of Rochester Center for Coherence and Quantum Science, Rochester, New York 14627, USA}

\begin{abstract}
Feedback control in open quantum dynamics is crucial for the advancement of various coherent platforms. 
However, currently only a handful of feedback master equations exist in the literature, which are restricted to specific types of feedback.
In this letter we first introduce a unifying framework,  based on a single general equation, that describes all possible feedback schemes in sequentially (and continuously) measured systems, and from which all previous results follow.
Next, we specialize it to the case of quantum jumps and introduce a new type of feedback based on the channel of the last detected jump, as well as the time elapsed since it occurred. 
Our description is experimentally grounded, and naturally allows for the introduction of realistic effects, such as time-delays in the feedback loop. 
We illustrate our results with two time-dependent feedback protocols conditioned on quantum-jump detections: one achieving population inversion of a two-level system against a thermal bath, and another enabling real-time reversal of quantum transitions, both admitting steady-state solutions.
\end{abstract}

\maketitle

\begin{figure}
    \centering
      \includegraphics[scale=0.43]{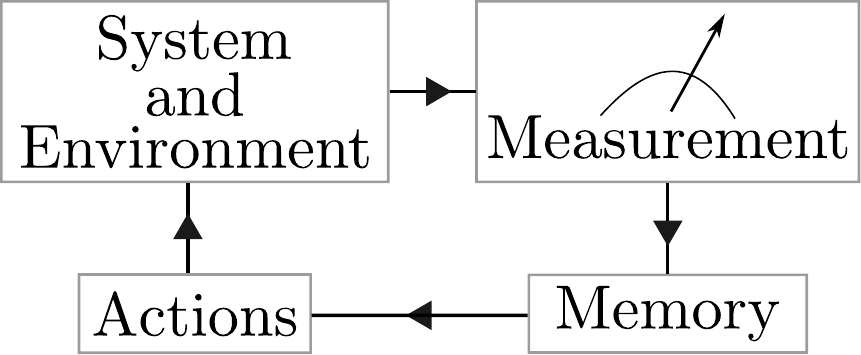}
    \caption{General feedback diagram. A detection outcome is stored in memory and used in a feedback action that can modify the system dynamics (e.g., the Hamiltonian $H$), the environment (e.g., bath temperature), or the measurement process itself.}
    \label{fig: FB diagram and summary of results}
\end{figure}

\textit{\textbf{Introduction}}.---
Feedback refers to the process of dynamically controlling a system based on previous detection outcomes (see Fig.~\ref{fig: FB diagram and summary of results}). 
It plays a central role across quantum coherent platforms, including cooling protocols \cite{PhysRevB.68.235328,PhysRevLett.90.043001,PhysRevA.74.012322,PhysRevLett.96.043003,PhysRevA.91.043812,PhysRevLett.117.163601,PhysRevLett.123.223602,PhysRevLett.122.070603,PhysRevA.107.023516,CoolingGui}, quantum error correction \cite{QECwithFeedback,QuantumThermodynamicsForQuantumComputing}, entanglement generation \cite{Riste2013}, thermodynamics \cite{Pekola2015,PatrickFluctuationTheo,DiscreteFeedbackExp}, transport \cite{RevModPhys.75.1}, control \cite{Vijay2012,DiscreteFeedback2,DiscreteFeedback3}, and quantum batteries \cite{Mitchison2021chargingquantum}. 
It has enabled experimental realizations of Maxwell’s demon in the quantum regime \cite{QuantumDemon1,QuantumDemon2,QuantumDemon3}, motivating generalized formulations of the second law that incorporate feedback \cite{SecondLawFeedback1,SecondLawFeedback2,SecondLawFeedback3} and underscoring its importance in both theory and experiment.

The theoretical description of feedback in open quantum systems is non-trivial. 
Only a limited set of feedback master equations has been derived~\cite{QECwithFeedback,10.21468/SciPostPhys.17.3.083,PatrickQFPME,WisemanMilburn1993_Homodyne,Wiseman1994_Feedback,FPTLandi}, typically restricted to specific measurements and feedback schemes. 
For example, Ref.~\cite{PatrickQFPME} considers weak Gaussian measurements~\cite{weakMeasurements1} with low-pass filtering, while Ref.~\cite{WisemanMilburn1993_Homodyne} employs feedback based on homodyne outcomes.
Consequently, realistic feedback protocols often rely on stochastic differential equations~\cite{WisemanBook,Belavkin1983_Control,Belavkin1992_Filtering,Belavkin1987_NonDemolition,Belavkin1992_Collapse,Korotkov2001_Qubit,Jacobs2014_Book,Zhang2017_Review}, which are noisy,
computationally demanding, and lacking in physical insights.
In contrast, feedback master equations often yield analytical results valid across broad parameter ranges, revealing parameter dependencies and universal features.

The lack of feedback equations is dire for protocols based on quantum jumps, which are central to many experiments, from optical cavities~\cite{Sayrin2011RealTimeFeedback,Minev2019CatchingReverseQuantumJump} to quantum dots~\cite{PhysRevLett.117.206803,PhysRevApplied.23.044063}, where jumps can be detected in real time. 
One of the few available results is Ref.~\cite{Wiseman1994_Feedback}, which implements the following protocol: upon detection of a quantum jump, an instantaneous quantum channel (e.g., a pulse) is applied.
However, this framework does not capture many relevant scenarios, such as applying a coherent drive for a finite duration or transiently modifying system parameters after a jump.
These protocols have been implemented in quantum-dot experiments, where jumps are monitored via a quantum point contact~\cite{PhysRevLett.117.206803,PhysRevApplied.23.044063}, although a corresponding feedback equation is still unavailable~\footnote{It turns out that if the system is incoherent, this example can be described using classical master equations~\cite{PhysRevB.101.165404}, but this is no longer possible in a genuinely coherent scenario.}.

In this Letter, we fill this gap by introducing a general framework for sequential feedback protocols that provides analytical treatments in previously inaccessible regimes, all derived from a single equation (Result~\ref{result: 1}).
In the Supplemental Material~\cite{SupMat}, we recover all previouslly known feedback equations as special cases of Result~\ref{result: 1}.
As our second main result, we specialize this framework  to quantum jumps and derive a new class of feedback that is based on both the channel of the last jump, as well as the time elapsed since the last jump (Result~\ref{result: 2}). This is motivated by recent experiments, such as in quantum dot devices.

We apply our framework to two time-dependent feedback protocols conditioned on quantum-jump detections: (i) population inversion of a qubit coupled to a thermal bath, and (ii) real-time reversal of quantum transitions. These experimentally relevant schemes~\cite{Sayrin2011RealTimeFeedback,Minev2019CatchingReverseQuantumJump,PhysRevLett.117.206803,PhysRevApplied.23.044063} lie beyond the scope of previous deterministic feedback formulations~\cite{Wiseman1994_Feedback,FPTLandi,PatrickQFPME,WisemanMilburn1993_Homodyne}.
Further developments and applications of this framework are presented in follow-up works~\cite{Rosal2025MemoryStatFB,Rosal2025FCSwithFB}.

\begin{figure}
    \centering
    \includegraphics[width=\linewidth]{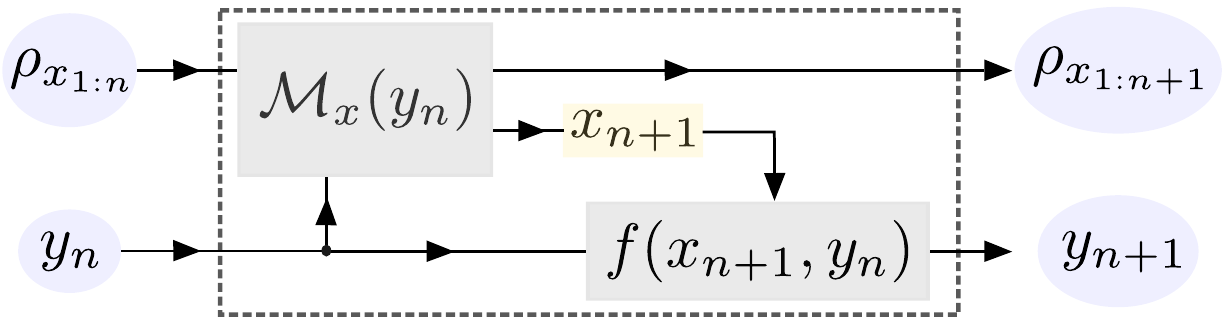}
    \caption{Conditional evolution under feedback. 
    The instruments $\mathcal{M}_x$ describe both the system dynamics and the measurement process at each step. 
    The quantities $\rho_{x_{1:n}}$ and $y_n$ denote, respectively, the system state and memory after $n$ detections given the dataset $(x_1, \dots, x_n)$.
    Feedback is implemented by using the memory $y_n$ to modify the instrument, while the measurement yields the outcome $x_{n+1}$ and updates the memory as $y_{n+1} = f(x_{n+1}, y_n)$.}
    \label{fig: stochastic evolution}
\end{figure}

{\bf \emph{Memory and data processing.}}$-$
\label{sec:seq_measurements}
Feedback can be described in terms of the diagram in  Fig.~\ref{fig: FB diagram and summary of results}. 
A system is sequentially measured, yielding a string of random outcomes $(x_1,x_2,\ldots,x_n,\ldots)$, where $x_i$ is the outcome at the discrete time $t_i$ (the continuous-time version can be obtained as a limit~\footnote{The continuous monitoring limit is obtained by applying measurements at regular intervals $t_i = i\delta t$ with $\delta t > 0$, and then taking the limit $\delta t \to 0$.}).
The feedback protocol consists of a set of possible actions, which can be applied to the system, and which are chosen based on the observed data. 
This could be, e.g., applying a laser drive, or changing a gate voltage.
The key aspects that define the protocol are (i) the type of measurement (projective, diffusive, quantum jumps, etc.); and (ii) what data is actually being used to construct the feedback actions. 
The latter is particularly non-trivial. At step $n$, the data available is in principle the full set $x_{1:n}\equiv(x_1,x_2,\ldots,x_n)$, but not all of it might actually be needed.

A general approach stores part or all of the past measurement record in a memory, denoted here by $y_n$. Hence, the memory is a function of the previous outcomes, $y_n = y_n(x_{1:n})$, and may represent either the full dataset, $y_n = (x_1,x_2,\ldots,x_n)$, or a compressed version of it.
Adding a memory requires an additional data processing step, where the new  value $y_n$ is updated according to some function $y_n = f_n(x_n,y_{n-1})$, that depends on the latest outcome $x_n$, as well as the past memory value $y_{n-1}$.
With an appropriate choice of data-processing function $f_n$, one can describe all possible feedback schemes.

For instance, the simplest feedback scheme uses only the latest outcome $x_n$ \cite{Wiseman1994_Feedback,WisemanMilburn1993_Homodyne}, with the data-processing function $f_n(x_n,y_{n-1}) = x_n$, corresponding to a compression that discards all previous outcomes.
At the opposite extreme, one may retain the entire record $(x_1,\ldots,x_n)$ using $f_n(x_n,y_{n-1}) = \text{append}(y_{n-1},x_n)$, where each new outcome is appended to the memory, and no compression is applied. 
Furthermore, feedback schemes can depend on summary statistics of the record $(x_1,\ldots,x_n)$, such as the sample mean~\cite{FPTLandi}, $y_n = (x_1+\ldots+x_n)/n$, with the update function $f_n(x_n,y_{n-1}) = (x_n+ (n-1)y_{n-1})/n $.
Additional examples of data-processing functions are given in~\cite{SupMat}.

{\bf \emph{Sequential measurements.}}$-$
A general description of sequentially measured systems is given in terms of \emph{instruments} $\{\mathcal{M}_x\}$, i.e., trace non-increasing maps associated with outcome $x$, such that $\sum_{x}\mathcal{M}_x$ defines a quantum channel~\cite{PRXQuantum.2.030201}.
For a state $\rho$, the probability of outcome $x$ is $\mathrm{Tr}(\mathcal{M}_x \rho)$, and the post-detection state is $\mathcal{M}_x\rho/\mathrm{Tr}(\mathcal{M}_x\rho)$.
Instruments capture both the system dynamics and the measurement process, providing a unified framework for phenomena ranging from projective measurements to quantum jump detections.
For example, consider a system measured at regular intervals $\delta t>0$ with Kraus operators $\{V_x\}$, and intermediate dynamic evolution given by $\rho(t+\delta t)=\Lambda_{\delta t}\rho(t)$.
The corresponding instrument is $\mathcal{M}_x\rho = V_x\,[\Lambda_{\delta t}(\rho)]\,V_x^\dagger$.

The most general feedback mechanism consists of dynamically modifying the instrument $\mathcal{M}_{x_{n+1}}$ at each time step $t_{n+1}$, based on the memory $y_n$ stored at that step (see Fig.~\ref{fig: stochastic evolution}).
The stochastic dynamics in the presence of feedback is therefore described by the following rules
\begin{equation}
\begin{aligned}
    P(x_{n+1} | x_{1:n}) &= \mathrm{Tr}[\mathcal{M}_{x_{n+1}}(y_{n})\rho_{x_{1:n}}],
    \\[0.2cm]
    \rho_{x_{1:n+1}} &= \frac{\mathcal{M}_{x_{n+1}}(y_{n})\rho_{x_{1:n}}}{P(x_{n+1} | x_{1:n})},
\end{aligned}    
\end{equation}
where $\rho_{x_{1:n}}$ denotes the state conditioned on the dataset $x_{1:n}$.

Our goal is to derive a deterministic equation governing the time evolution of a general feedback protocol.
To do that we define the memory-resolved state \cite{FPTLandi,PatrickQFPME} (also called hybrid state~\cite{10.21468/SciPostPhys.17.3.083,Layton2024healthiersemi,PhysRevA.107.062206,Oppenheim2023objective,PhysRevX.13.041040}) as $\varrho_n(y) = E[\rho_{x_{1:n}} \delta_{y, y_n}]$, where $E[\cdot]$ denotes the  average over trajectories, and $\delta_{y, y_n}$ is the Kronecker delta. 
The trace of this state gives the distribution of the memory function $y_n$, $\mathrm{Tr}[\varrho_n(y)] = P(y_n = y)$, while the marginalization over all realizations of $y_n$ yields the unconditional state of the system, $\sum_y \varrho_n(y) = \bar{\rho}_n$.
Notice that $\rho_{x_{1:n}}$ is a stochastic (conditional) state, but $\varrho_n(y)$ is deterministic.
Our first main result is:
\begin{result}
\label{result: 1}
If an instrument $\{\mathcal{M}_x(y)\}$ depends only on a causal memory function $y_n = f_n(x_n, y_{n-1})$, the corresponding memory-resolved state $\varrho_n(y)$ evolves according to the deterministic equation
\begin{equation}
    \label{eq: deterministic eq 1}
    \varrho_{n+1}(y) = \sum_{x',y'} \delta_{y, f_{n+1}(x',y')} \mathcal{M}_{x'}(y') \varrho_n(y')~,
\end{equation}
where the sum runs over all possible outcomes $x'$ and all possible realizations $y'$ of $y_n$.
\end{result}

The proof can be found in~\cite{SupMat}.
Since $\varrho_n(y)$ is resolved in $y$, the state at step $n+1$ should be just a sum over all possible  trajectories  consistent with the constraint $y_n = f_n(x_n, y_{n-1})$.
Result~\ref{result: 1} provides a single, general equation for any feedback scheme in sequentially measured systems, and all previously known feedback equations~\cite{PatrickQFPME, WisemanMilburn1993_Homodyne, Wiseman1994_Feedback, FPTLandi} are recovered as special cases~\cite{SupMat}.
In the remainder of the paper, we employ this framework to derive a new class of feedback strategies based on quantum jump detections.

{\bf \emph{Quantum jumps.}}$-$
We consider a system that is described by the quantum master equation
\begin{equation}\label{QME}
 \partial_t \rho_t =\mathcal{L}\rho= -i \big[H,\rho_t\big] +  \sum_{k} L_k^{} \rho_t L_k^\dagger - \frac{1}{2} \big\{ L_k^\dagger L_k^{},\rho_t\big\},
\end{equation}
with Hamiltonian $H$ and jump operators $L_k$. 
We assume that there is a subset of quantum jumps $k \in \Sigma$, which can be experimentally monitored.
Within the quantum jump unravelling, this means that the dynamics for an infinitesimal step $\delta t$ can be decomposed into a set of instruments~\cite{Tutorial}
\begin{align}
    \label{eq: no-jump instrument}
    \mathcal{M}_0\rho &= (1+\delta t \mathcal{L}_0 ) \rho, 
    \\[0.2cm]
    \label{eq: jump instrument}
    \mathcal{M}_k\rho &= \delta t \mathcal{J}_k \rho,
\end{align}
where $\mathcal{J}_k\rho = L_k^{} \rho L_k^\dagger$, $\mathcal{L}_0  = \mathcal{L} - \sum_{k\in\Sigma} \mathcal{J}_k$.
At the stochastic level, the outcome at step $t_n = n \delta t$ is either $x_n = 0$ (no jump) or $x_n = k$ for a jump of type $k$.
Our goal is to introduce feedback schemes which allow for both the Hamiltonian $H$, and the jump operators, to be modified in real time based on these measurements.

We consider two types of memories.
First, one that records the channel $k\in \Sigma$ of the last jump. This can be constructed with a function $f_n$ of the form
\begin{equation}
\label{eq: jump_filter}
k_n = x_n + k_{n-1}\delta_{x_n, 0} ~.
\end{equation}
Second, a memory that records how much time has elapsed since the last jump, which is built using
\begin{equation}
    \label{eq: counting_filter}
    \tau_{n} = \delta_{x_n,0} (\tau_{n-1}+\delta t)~.
\end{equation}
We refer to them as the jump and counting memories, respectively.
A general feedback mechanism based on both jump and counting memories is defined by the total memory function $y_n = (k_n,\tau_n)$, with the instruments given in Eqs.~\eqref{eq: no-jump instrument} and \eqref{eq: jump instrument}.  
In this setting, the feedback allows both the system Hamiltonian and the jump operators at time $t_{n+1}$ to depend on the last detected jump $k_n$ and the elapsed time $\tau_n$, taking the forms $H(k_n,\tau_n)$ and $L_k(k_n,\tau_n)$, which in turn define the corresponding super-operators $\mathcal{L}_0(k_n,\tau_n)$ and $\mathcal{J}_k(k_n,\tau_n)$.
Starting from Result~\ref{result: 1}, we show in~\cite{SupMat} the following result:
\begin{result}
    \label{result: 2}
    Let $\varrho_t(k,\tau)$ denote the state of the system resolved in both the jump and counting memory functions~\eqref{eq: jump_filter}, \eqref{eq: counting_filter}.
    In the limit $\delta t\to 0$ this evolves according to the deterministic equations
    \begin{eqnarray}
    \label{eq: deterministic equations combined signal1}
        \varrho_t(k,0) &=& 2 \delta(t) \delta_{k,\bar{k}}\bar{\rho}_0 + \sum_{q\in\Sigma}\int\limits_0^t~d\tau \mathcal{J}_k(q,\tau) \varrho_t(q,\tau)
        ,\\
    \label{eq: deterministic equations combined signal2}
        \varrho_t(k,\tau) &=& G(k,\tau) \varrho_{t-\tau}(k,0),
    \end{eqnarray}
    where $G(k,\tau)\equiv\mathcal{T}\left[e^{\int_0^\tau ds \mathcal{L}_0(k,s)}\right]$ is the propagator of the no-jump evolution and $\mathcal{T}[\cdot]$ is the time-ordering operator.
\end{result}

In the limit $\delta t \to 0$, the memory functions defined in Eqs.~\eqref{eq: jump_filter} and \eqref{eq: counting_filter} become continuous-time stochastic processes, $k_t$ and $\tau_t$, specifying that at time $t$ the last jump occurred in channel $k_t$ at time $t-\tau_t$.
The quantity $\mathrm{Tr}[\varrho_t(k,\tau)]$ is the joint distribution of $(k_t,\tau_t)$, and from~\eqref{eq: counting_filter} it follows that $\tau_t \in [0,t]$. 
The resolved state is normalized as $\sum_{k\in \Sigma}\int_0^t \mathrm{Tr}[\varrho_t(k,\tau)]d\tau = 1$, and the unconditional state is recovered as 
\begin{equation}\label{unconditional_state_jump}
    \bar{\rho}_t = \sum_{k \in \Sigma}\int_0^t d\tau \varrho_t(k,\tau)~.
\end{equation}
The term $2\delta(t)\delta_{k,\bar{k}}\bar{\rho}_0$ encodes the initial condition, with system state $\bar{\rho}_0$ and memory values $k_0=\bar{k}$ and $\tau_0=0$. We adopt the convention $\int_0^t d\tau\, \delta(t-\tau)=1/2$.
Result~\ref{result: 2} enables time-dependent control actions, such as temporarily modifying the system Hamiltonian or environmental parameters. 
In contrast, previous jump-based schemes~\cite{Wiseman1994_Feedback} relied solely on the current jump detection.
Although Result~\ref{result: 2} cannot, in general, be expressed as a differential equation, it remains amenable to analytical treatment -- as detailed in~\cite{SupMat} -- especially in the stationary regime.

Let us consider the special case where the feedback depends only on the last detected jump recorded by $k_t$, regardless of the time elapsed since its detection.
We show in~\cite{SupMat} that the marginal state $\varrho_t(k) = \int_0^t d\tau \varrho_t(k,\tau)$ will evolve as
\begin{eqnarray}
\label{eq: Patrick's equation}
    \partial_t \varrho_t(k) &=& -i \big[H(k),\varrho_t(k)\big] - \frac{1}{2} \sum_{k' \in \Sigma} \big\{ L_{k'}^\dagger(k)L_{k'}(k),\varrho_t(k)\big\} \nonumber \\
                             & &  +\sum_{k' \in \Sigma}\ L_k^{}(k') \varrho_t(k') L_k^\dagger(k'). 
\end{eqnarray}
Equation~\eqref{eq: Patrick's equation} represents a system of coupled equations, one for each jump channel $k \in \Sigma$, describing protocols that update the Hamiltonian and/or jump operators based on $k_t$. 
A similar equation was derived in~\cite{Blanchard1995EEQT} in a different context.

Returning  to the general Result~\ref{result: 2}, the situation simplifies when the feedback is applied only on the no-jump part of the dynamics (e.g. in the Hamiltonian). 
In this case $\mathcal{J}_k(k',\tau')= \mathcal{J}_k$ and 
Eqs.~\eqref{eq: deterministic equations combined signal1}-~\eqref{unconditional_state_jump}  imply that $ \varrho_t(k,\tau) = 2 \delta(t-\tau)\delta_{k,\bar{k}}G(k,\tau)\bar{\rho}_0+G(k,\tau)\mathcal{J}_k\bar{\rho}_{t-\tau}$. 
The memory-resolved state can hence be written in terms of the unconditional state $\bar{\rho}_t$ which, in turn, satisfies the closed integral equation
\begin{equation}
    \label{eq: evolution of the unconditional state for Result 2}
    \bar{\rho}_t = G(\bar{k},t) \bar{\rho}_0 + \sum_{k\in\Sigma}\int_0^td\tau 
    G(k,\tau)
    \mathcal{J}_k \bar{\rho}_{t-\tau}~.
\end{equation}
The steady-state, if it exists, is obtained by setting $t\to \infty$.
In this case, $\lim_{t\to\infty} G(k,t)=0$~\cite{SupMat}, so $\bar{\rho}_{\rm ss}$ will therefore be the solution of the algebraic equation

\begin{equation}
\label{eq: unconditional steady state for Result 2}
    \bar{\rho}_{\rm ss} = \Omega \bar{\rho}_{\rm ss},
    \qquad 
    \Omega = \sum_{k\in\Sigma}\int_0^\infty d\tau 
    G(k,\tau)
    \mathcal{J}_k,
\end{equation}
which also provides the steady-state memory-resolved state $\varrho_{\text{ss}}(k,\tau) = G(k,\tau)\mathcal{J}_k\bar{\rho}_{\text{ss}} $. 
In turn, this can be used to calculate various time-related properties, such as waiting-time distributions, the average time between jumps, etc. The ability to describe jump statistics under a general feedback protocol, as given by $\mathrm{Tr}[\varrho_t(k,\tau)]$, is a novel feature not captured by any previous feedback formulation.

\begin{figure}
    \centering
     \includegraphics[width=\linewidth]{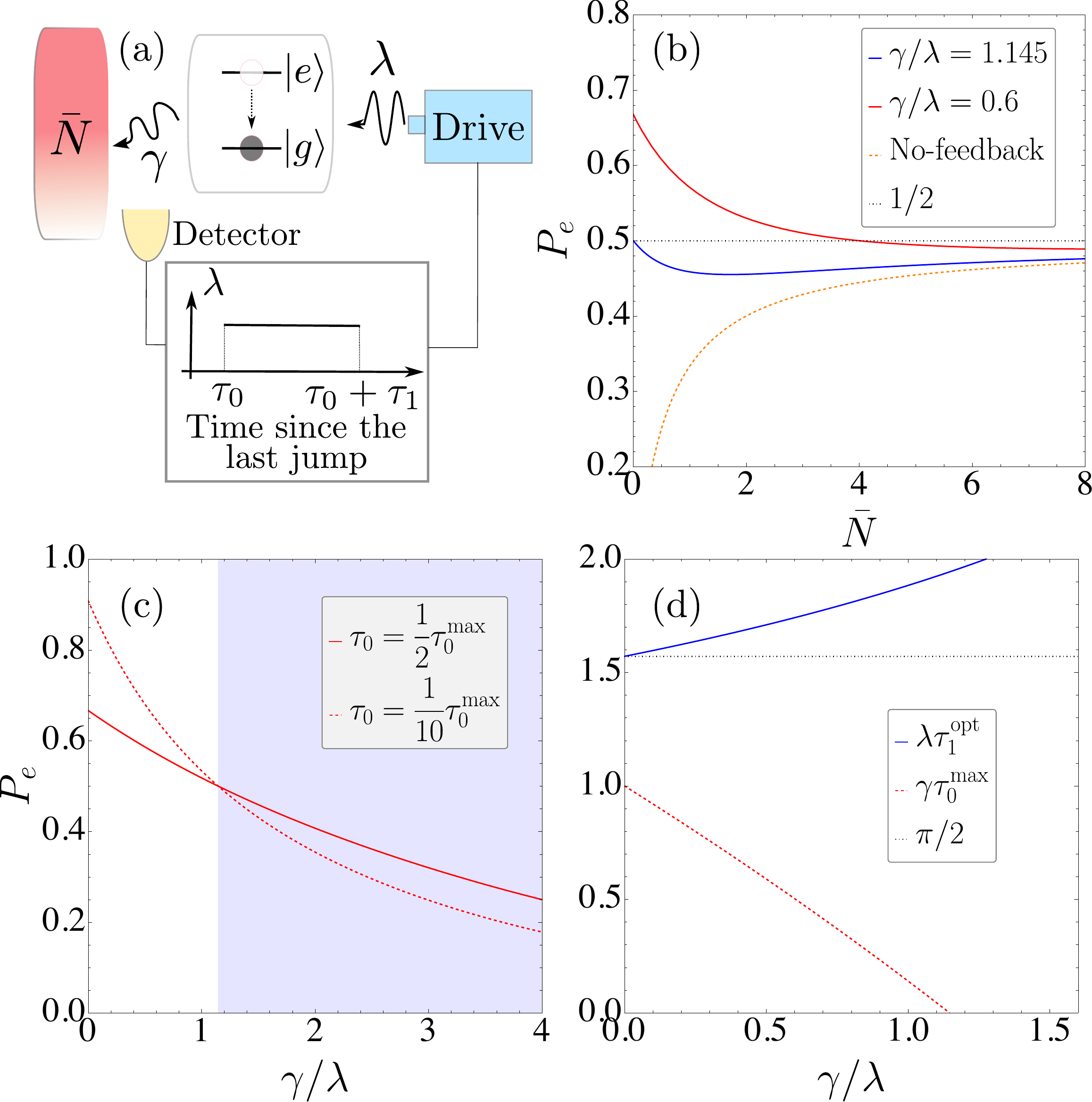}
     \caption{
    Population inversion via time-dependent feedback with quantum jumps.
    { (a)} Upon detection of a decay event, an external drive is applied to return the qubit to the excited state. 
    { (b)} Population of the excited state without feedback delay ($\tau_0 = 0$), and considering the optimal drive duration $\tau_1^{\text{opt}}$ (Eq.~\eqref{eq: results of the example - optimal time}). The no-feedback line corresponds to a qubit coupled to a thermal bath without an external drive. 
    {(c)} Effect of feedback delay on the excited-state population for $\bar{N}=0$. 
    {(d)} Maximum allowable feedback delay and the optimal drive duration for $\bar{N}=0$.   }
     \label{fig: plots}
\end{figure}

\textit{\textbf{Example 1: inversion protocol.}}-- 
Two-level systems (qubits) in excited states act as single-photon sources of non-classical light~\cite{Hofheinz2008GenerationFock} and as key elements in quantum computing~\cite{Maring2024VersatileSinglePhotonPlatform}.
Coupled to a thermal bath, they relax to a steady state with a higher ground-state population than excited-state population ($P_g>P_e$).
Our goal is to achieve population inversion ($P_g<P_e$) via time-dependent feedback and continuous monitoring of quantum jumps.
Let $\ket{e}$ and $\ket{g}$ denote the excited and ground states with energy gap $\omega$. 
The thermal reservoir has temperature $T$, and the jump operators are
$L_{-} = \sqrt{\gamma(\bar{N}+1)} ~|g\rangle\langle e|$
and 
$L_+ = \sqrt{\gamma\bar{N}}~|e\rangle\langle g|$ describing, respectively, the emission and absorption of a quanta, where $\gamma>0$ is the coupling strength, and $\bar{N} = (\text{exp}(\omega/T)-1)^{-1}$.

The protocol is defined as follows (see Fig.~\ref{fig: plots}(a)). 
In the rotating frame at frequency $\omega_d$, the qubit Hamiltonian reads $H = -\frac{\Delta}{2}\sigma_z$, where $\sigma_{x,y,z}$ are the Pauli matrices and $\Delta \equiv \omega - \omega_d$ (see~\cite{SupMat} for technical details). 
Upon detecting a thermal emission ($\ket{e}\!\to\!\ket{g}$) and observing no subsequent detection for a time $\tau_0$, an external drive with amplitude $\lambda$ and frequency $\omega_d$ is applied for a duration $\tau_1$, and the Hamiltonian becomes $H = -\frac{\Delta}{2}\sigma_z + \lambda\sigma_x$. 
In essence, after an emission and a time $\tau_0$, the drive induces Rabi oscillations for a time $\tau_1$, transferring population from $\ket{g}$ back to $\ket{e}$.

This protocol is described by Result~\ref{result: 2}, as it depends on both the last detected jump and the elapsed time.
The End Matter provides analytical steady-state results, including the optimal drive duration $\tau_1^{\text{opt}}$ [Eq.~\eqref{eq: results of the example - optimal time}] that maximizes the excited-state population. 
Notably, $\tau_1^{\text{opt}}$ generally differs from a $\pi$-pulse, as dissipation requires a longer drive, highlighting the interplay between coherence and dissipation in the feedback dynamics. 
Determining this optimal time from stochastic simulations would be cumbersome due to the large parameter space.
Fig.~\ref{fig: plots}(b) shows $P_e$ as a function of $\bar{N}$ for the ideal case without time delay ($\tau_0 = 0$) and considering the optimal drive duration $\tau_1^\text{opt}$. We find that for $\gamma/\lambda \leq 1.145$, there exists a critical value $\bar{N}_{\text{c}} \geq 0$ such that
$P_e>1/2$ for all $\bar{N} < \bar{N}_{\text{c}}$.
This therefore establishes the population inversion threshold.
In Fig.~\ref{fig: plots}(c), we illustrate how  $P_e$ is affected by varying the delay $\tau_0$ for $\bar{N} = 0$ [Eq.~\eqref{eq: results of the example - population for zero temp}], demonstrating how increasing $\tau_0$ deteriorates the population inversion. 

Fig.~\ref{fig: plots}(d) shows the optimal drive duration $\tau_1^{\text{opt}}$ [Eq.~\eqref{eq: results of the example - optimal time}] and the maximum delay $\tau_0^{\text{max}}$ [Eq.~\eqref{eq: results of the example - maximum delay}] enabling population inversion for $\bar{N}=0$.
In the strong-drive regime ($\gamma \ll \lambda$), we find that $\tau_0^{\text{max}} \sim 1/\gamma$, indicating that the maximum delay is on the order of the system's natural timescale $\gamma^{-1}$. Conversely, $\tau_0^{\text{max}} = 0$ for $\gamma/\lambda > 1.145$, reflecting that population inversion is no longer possible in this regime, regardless of the feedback delay. 
Additionally, we observe that $\gamma\tau_1^{\text{opt}}$ increases with $\gamma/\lambda$, implying that as the drive strength $\lambda$ decreases relative to $\gamma$, the drive must remain on for a longer duration to optimize $P_e$. On the other hand, when dissipation is weak ($\gamma \ll \lambda$), the optimal drive duration recovers the $\pi$-pulse.

\begin{figure}
    \centering
     \includegraphics[width=\linewidth]{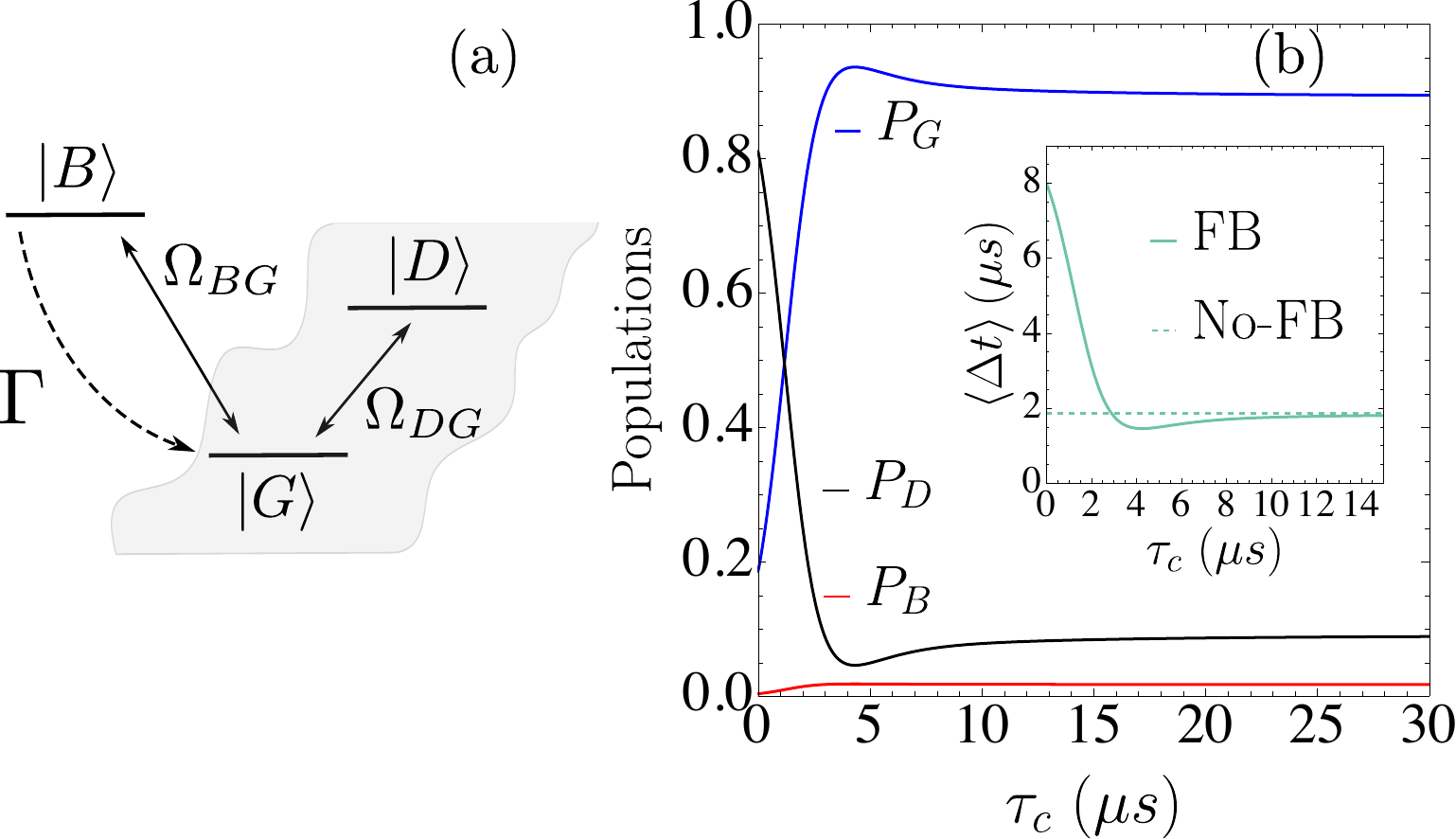}
    \caption{Reverting quantum transitions.
    {(a)} Three-level system with a hidden transition~\cite{Minev2019CatchingReverseQuantumJump} between the ground $\ket{G}$ and dark $\ket{D}$ states (shaded region).
    The states $\ket{B}$ and $\ket{G}$ are coupled to an effective thermal bath, analogous to the qubit setup in Example~1, with coupling rate $\gamma_B$ and Bose–Einstein occupation $\bar{N}_B$, so that $\Gamma = \gamma_B(\bar{N}_B + 1)$. The protected states $\ket{D}$ and $\ket{G}$ are only weakly coupled to a second bath characterized by $\gamma_D$ and $\bar{N}_D$.
    {(b)} Populations of the feedback steady state as a function of the waiting time $\tau_c$. The inset shows the average time $\braket{\Delta t}$ between jumps $\ket{B}\!\to\!\ket{G}$ in the stationary regime. The limit $\tau_c \!\to\! \infty$ corresponds to the no-feedback case, where no rotations are applied. 
    Parameters: $\Omega_{BG}/(2\pi) = 0.9$ MHz, $\gamma_B = 10~ \Omega_{BG} $, $\Omega_{DG}/(2\pi) = 14$ kHz, $\gamma_D = 0.875~\Omega_{DG}$, $\bar{N}_B = 0.01$, $\bar{N}_D = 0.05$, $\theta = \phi = \pi/2$.
   }
    \label{fig:nature-example}
\end{figure}

\textit{\textbf{Example 2: reverting quantum transitions.}}-- 
Ref.~\cite{Minev2019CatchingReverseQuantumJump} showed that quantum transitions can be continuously monitored and reversed in real time via feedback.
There, upon detecting the beginning of a transition, a pulse is applied after a time $\tau_c$ to reverse its evolution, originally described using stochastic simulations.
We apply our formalism to this setup with photon-counting detection~\cite{PhysRevLett.57.1699,PhysRevLett.57.1696,PhysRevLett.56.2797,PhysRevLett.54.1023}.
Rather than averaging many trajectories, our method yields the steady state from a single eigenvector equation and directly provides stationary jump statistics, including the mean inter-jump time.

The system [Fig.~\ref{fig:nature-example}(a)] consists of a ground state $\ket{G}$, a protected (dark) state $\ket{D}$ designed to minimally couple to any dissipative environment or measurement apparatus, and a bright state $\ket{B}$ that decays incoherently to $\ket{G}$ with coupling rate $\Gamma$, emitting a photon.
A weak Rabi drive $\Omega_{DG}$ couples $\ket{G}$ and $\ket{D}$, while a stronger drive $\Omega_{BG}$ satisfies $\Omega_{DG} \ll \Omega_{BG} \ll \Gamma$, linking the populations of $\ket{G}$ and $\ket{B}$.
The incoherent jump $\ket{B}\!\to\!\ket{G}$ is monitored by detecting the emitted photon, which signals occupation of $\ket{G}$, while a prolonged absence of photons indicates coherent evolution within the $\{\ket{G},\ket{D}\}$ subspace.
The protocol proceeds as follows: after detecting the incoherent jump $\ket{B}\!\to\!\ket{G}$ and observing no further detection for a time $\tau_c$, a pulse is applied on the $\{\ket{G},\ket{D}\}$ Bloch sphere, corresponding to a rotation by angle $\theta$ around the axis $\hat{n} = (\cos\phi,\sin\phi,0)$, enabling reversal or completion of the $\ket{G}\!\leftrightarrow\!\ket{D}$ transition.

Fig.~\ref{fig:nature-example}(b) shows the steady-state populations as a function of the waiting time $\tau_c$ between the last detected jump and the application of a $(\pi/2)$--pulse around the $y$--axis ($\phi=\theta=\pi/2$).
For short $\tau_c$, the system remains mostly in $\ket{G}$ after the jump $\ket{B}\!\to\!\ket{G}$, so the $(\pi/2)$--pulse transfers population to $\ket{D}$, increasing its steady-state occupation.
However, the system may evolve toward $\ket{D}$ before the pulse, which then drives it back to $\ket{G}$, maximizing the $\ket{G}$ population.
The inset shows the average interval $\braket{\Delta t}$ between photon detections (jumps $\ket{B}\!\to\!\ket{G}$) as a function of $\tau_c$. Maximizing $P_G$ enhances coherent excitation to $\ket{B}$, making jumps $\ket{B}\!\to\!\ket{G}$ more frequent and reducing $\braket{\Delta t}$, while increasing $P_D$ suppresses excitations and lengthens $\braket{\Delta t}$. 
This illustrates how feedback directly shapes the jump statistics, which can also be analyzed in terms of the pulse angles $\theta$ and $\phi$.

\textit{\textbf{Conclusion}}.---
We derived a deterministic equation that describes a general feedback protocol, encompassing both discrete and continuous feedback. 
It recovers previous results as particular cases, and allows for the derivation of novel feedback schemes, such as time-dependent strategies. 
Our framework offers numerical efficiency and conceptual clarity through closed-form expressions, allowing analytical identification of operating regimes and critical thresholds -- advantages crucial for theory and experiment.


\textit{\textbf{Acknowledgments}}.--- The authors acknowledge fruitful discussions with Mark Mitchison, Guilherme Fiusa, Abhaya Hegde.
P.P.P. acknowledges funding from the Swiss National Science Foundation (Eccellenza Professorial Fellowship PCEFP2\_194268).

\bibliography{feedback-refs}




\newpage

\onecolumngrid

\begin{center}
    \textbf{End Matter}
\end{center}

\twocolumngrid

\setcounter{section}{0}
\setcounter{equation}{0}
\setcounter{figure}{0}
\setcounter{table}{0}
\setcounter{page}{1}
\renewcommand{\theequation}{S\arabic{equation}}
\renewcommand{\thefigure}{S\arabic{figure}}

\begin{figure}
    \centering
     \includegraphics[width=\linewidth]{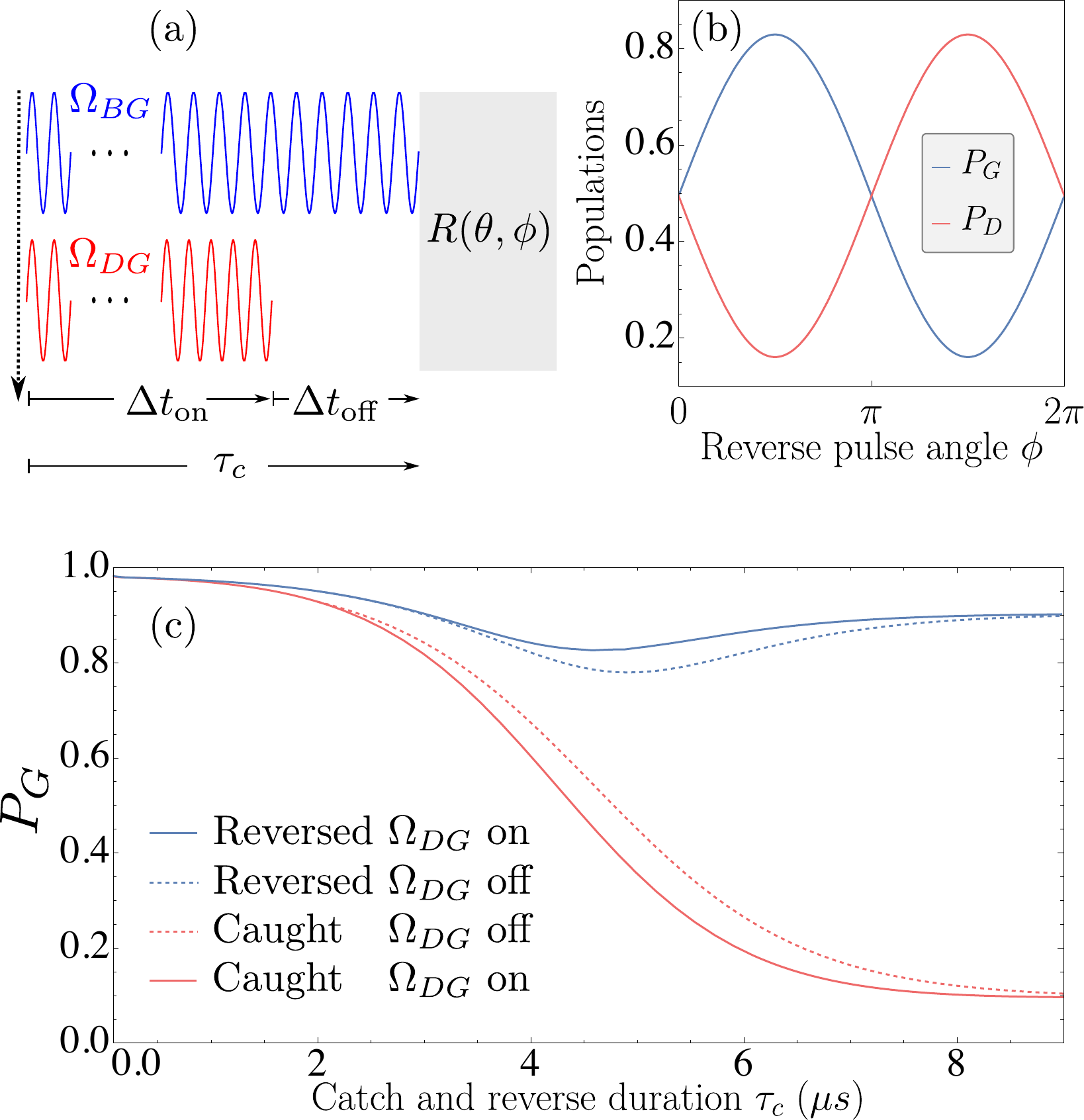}
     \caption{
    Populations of the system's state immediately after the pulse.  
    (a) The vertical arrow on the left-hand side represents the last detected jump $\ket{B}\to\ket{G}$, and the horizontal axis indicates the time elapsed since its detection.
    The system is driven by $\Omega_{BG}$ (blue line) and $\Omega_{DG}$ (red line). After a time $\tau_c$ has elapsed since the last jump, a pulse $R(\theta,\phi)$ is applied in the $\{\ket{G}, \ket{D}\}$ subspace. 
    The drive $\Omega_{DG}$ can either remain on during the full interval $\tau_c$ or be turned off after a time $\Delta t_{\text{on}}$.  
    (b) Populations of the ground $\ket{G}$ ($P_G$) and dark $\ket{D}$ ($P_D$) states after the pulse for $\theta = \pi/2$, as a function of the direction angle $\phi$, for $\tau_{c} = 4.2~\mu\text{s}$.  
    (c) Population $P_G$ of the ground state as a function of the waiting time $\tau_c$. Solid lines correspond to the case where the drive $\Omega_{DG}$ remains on during $\tau_c$, while dashed lines correspond to the case where $\Omega_{DG}$ is turned off after a time $\Delta t_{\text{on}}$, as illustrated in (a).
    Blue curves represent the feedback case (rotation applied), while red curves correspond to the no-feedback case (no rotation after $\tau_c$).
    In the feedback case, for each $\tau_c$ we determine the optimal pulse angles $\{\theta(\tau_c), \phi(\tau_c)\}$ that maximize $P_G$, and use them to compute $P_G(\tau_c)$.  
    The parameters $\gamma_{B,D}$, $\bar{N}_{B,D}$, $\Omega_{BG}$, and $\Omega_{DG}$ are the same as those used in Fig.~\ref{fig:nature-example}.
}
     \label{fig:nature-paper-condstate}
\end{figure}

\emph{Appendix: population inversion protocol ---} The feedback protocol considered in Example~1 affects only the system's Hamiltonian; hence, the steady state can be obtained from Eq.~\eqref{eq: unconditional steady state for Result 2}. 
A detailed derivation is provided in the Supplemental Material~\cite{SupMat}, which gives the analytical expression for the excited-state population $P_e = \langle e|\rho_{ss}|e\rangle$ at resonance ($\Delta = 0$) as a function of the bath parameters $\bar{N}$ and $\gamma$, and the feedback parameters $\lambda$, $\tau_0$, and $\tau_1$.
By maximizing $P_e$ with respect to $\tau_1$, one obtains the optimal duration $\tau_1^{\text{opt}}$ for which the external drive should remain on. We find that
\begin{equation}
\label{eq: results of the example - optimal time}
    \tau_1^{\text{opt}} = \frac{2\left[2\pi +\arctan{\left( \frac{p\sqrt{16-p^2}}{8-p^2} \right)}-\pi\theta(\sqrt{8}-p)\right]}{\lambda\sqrt{16-p^2}} ~,
\end{equation}
where $p = \gamma/\lambda$ and $\theta(x)$ is the Heaviside function. 
This is finite only for $p < 4$, and diverges otherwise.
It is noteworthy that $\tau_1^{\text{opt}}$ is independent of both the feedback delay $\tau_0$ and the bath occupation $\bar{N}$; it depends solely on the ratio $p = \gamma/\lambda$ between the coupling rate to the bath and the drive strength.

Next we consider a more realistic scenario where we have a time delay $\tau_0$ between the last detected jump and the moment the external drive is turned on. Even in this more general case, the optimal drive duration that maximizes the population of the excited state remains given by Eq.~\eqref{eq: results of the example - optimal time}~\cite{SupMat}. In the low-temperature limit $\bar{N} \rightarrow 0$, the expression for $P_e$, evaluated at the optimal drive duration, simplifies to
\begin{eqnarray}
\label{eq: results of the example - population for zero temp}
    \lim_{\bar{N}\rightarrow0}P_e = \frac{1}{2 - e^{-\gamma\tau_1^{\text{opt}}/2} +\frac{1}{4}\left(\frac{\gamma}{\lambda}\right)^2 + \gamma\tau_0}~.
\end{eqnarray}
In the limit $\bar{N} \!\to\! 0$, population inversion ($P_e > 1/2$) occurs only for time delays below a maximum value $\tau_0^{\text{max}}$, given by
\begin{eqnarray}
\label{eq: results of the example - maximum delay}
    \tau_0^{\text{max}} = \frac{4e^{-\gamma\tau_1^{\text{opt}}/2} -(\gamma/\lambda)^2}{4\gamma} ~.
\end{eqnarray}

It is worth mentioning that this protocol can be adapted to increase the population of the ground state: by turning on the drive when an absorption is detected, the drive will move the system toward the ground state through Rabi oscillations. In this case, the protocol effectively operates as a cooling protocol based on quantum jump detections.

\emph{Appendix: reverting quantum transitions ---} Ref.~\cite{Minev2019CatchingReverseQuantumJump} experimentally implemented the feedback protocol described in Example~2. Instead of directly monitoring the incoherent jump $\ket{B}\!\to\!\ket{G}$ via photon detection, they coupled the bright state $\ket{B}$ to a readout cavity. The population of $\ket{B}$ was then inferred by measuring the number of photons in the cavity, allowing them to determine whether the system was in $\ket{B}$ or in a superposition of $\ket{G}$ and $\ket{D}$.
However, as discussed in the Supplementary Material of Ref.~\cite{Minev2019CatchingReverseQuantumJump}, a photon-counting scheme (as adopted in Example~2) can serve as a first approximation to the experimental results obtained via indirect monitoring of $\ket{B}$ using homodyne detection in the cavity. The authors’ theoretical description employed stochastic equations accounting for the random homodyne signal, which were used to sample system trajectories conditioned on a given homodyne measurement.

A photon-counting scheme can be implemented by monitoring intermittent fluorescence from the bright state $\ket{B}$ in trapped ions~\cite{PhysRevLett.57.1699,PhysRevLett.57.1696,PhysRevLett.56.2797,PhysRevLett.54.1023}. This scheme tracks the incoherent jumps $\ket{B}\!\to\!\ket{G}$, and our formalism directly provides the steady state of this feedback protocol as the solution of an eigenvector problem, as well as the jump statistics in the stationary regime. 
The feedback protocol in Example~2 does not modify the jump operators; the feedback action consists of applying a pulse in the $\{\ket{G},\ket{D}\}$ subspace after a time $\tau_c$ has elapsed since the last detected jump. Consequently, the steady state is given by Eq.~\eqref{eq: unconditional steady state for Result 2} (see Supplemental Material~\cite{SupMat} for details).

Our framework provides the analytical solution of the system state immediately after the pulse, as reported in Fig.~4 of Ref.~\cite{Minev2019CatchingReverseQuantumJump}. The populations of this conditioned state, considering the photon-counting scheme of Example~2, are shown in Fig.~\ref{fig:nature-paper-condstate}. From our formalism (see~\cite{SupMat}), the state immediately after the pulse, assuming $\Omega_{DG}$ is continuously on, reads
\begin{equation}
\label{eq: nature-conditional-state1}
\rho_{\text{on}}(\tau_c) = \frac{R(\theta,\phi)~ e^{\tau_c \mathcal{L}_0^{(1)}}\left(\ket{G}\bra{G}\right)}{\mathrm{Tr}\left[R(\theta,\phi) ~e^{\tau_c \mathcal{L}_0^{(1)}}\left(\ket{G}\bra{G}\right)\right]},
\end{equation}
where $\mathcal{L}_0^{(1)}$ is the no-jump Liouvillian with both drives $\Omega_{DG}$ and $\Omega_{BG}$ active during $\tau_c$ (see Fig.~\ref{fig:nature-paper-condstate}(a)), and $R(\theta,\phi)$ represents the rotation in the $\{\ket{G},\ket{D}\}$ subspace.
In Example~2, we keep $\Omega_{DG}$ on continuously for simplicity, but we can also consider turning it off after a time $\Delta t_{\text{on}}$ since the last jump, keeping only $\Omega_{BG}$ active, as illustrated in Fig.~\ref{fig:nature-paper-condstate}(a). In this case, the state immediately after the pulse reads
\begin{equation}
\label{eq: nature-conditional-state2}
    \rho_{\text{off}}(\tau_c) = \frac{R(\theta,\phi)~ e^{\Delta t_{\text{off}}~ \mathcal{L}_0^{(2)}}e^{\Delta t_{\text{on}}~ \mathcal{L}_0^{(1)}}\left(\ket{G}\bra{G}\right)}{\mathrm{Tr}\left[R(\theta,\phi)~ e^{\Delta t_{\text{off}}~ \mathcal{L}_0^{(2)}}e^{\Delta t_{\text{on}}~ \mathcal{L}_0^{(1)}}\left(\ket{G}\bra{G}\right)\right]}~,
\end{equation}
where $\mathcal{L}_0^{(2)}$ is the no-jump Liouvillian with only $\Omega_{BG}$ active.

The interpretation of Eqs.~\eqref{eq: nature-conditional-state1} and \eqref{eq: nature-conditional-state2} is straightforward: since the jump $\ket{B}\!\to\!\ket{G}$ is a renewal process~\cite{Tutorial}, the system resets to $\ket{G}$ after each jump. If no jump occurs during the interval $\tau_c$, the system evolves under the appropriate no-jump Liouvillian, after which the rotation $R(\theta,\phi)$ is applied.
Both Eqs.~\eqref{eq: nature-conditional-state1} and \eqref{eq: nature-conditional-state2} follow directly from the memory-resolved steady state $\varrho_{ss}(k,\tau)$ in Result~\ref{result: 2}, representing the system state given a time $\tau_c$ since the last jump.
For Example~2, only the jump $\ket{B}\!\to\!\ket{G}$ is monitored, so the memory-resolved state depends solely on $\tau$, denoted by $\varrho_{ss}(\tau)$.

The memory-resolved steady state reads
\begin{equation}
    \varrho_{ss}(\tau) = \gamma_B(\bar{N}_B+1)\bra{B}\bar{\rho}_{ss}\ket{B}G(\tau)(\ket{G}\bra{G})~,
\end{equation}
where $G(\tau)$ is the no-jump propagator and $\bar{\rho}_{ss}$ is the unconditional steady state given by Eq.~\eqref{eq: unconditional steady state for Result 2}.  
For $\Omega_{DG}$ always on, $G(\tau)$ is
\begin{equation}
    G(\tau) =   \begin{cases}
                    e^{\tau \mathcal{L}_0^{(1)}}, & \text{if } \tau < \tau_c, \\
                  e^{(\tau-\tau_c) \mathcal{L}_0^{(1)}}R(\theta,\phi) e^{\tau_c \mathcal{L}_0^{(1)}} ,  & \text{if } \tau \ge \tau_c.
                \end{cases}
\end{equation}
 The trace $\mathrm{Tr}[\varrho_{ss}(\tau)]$ gives the stationary probability density of detecting a jump and observing no-jump evolution for a time $\tau$.
 This provides the jump statistics under feedback, showing how the protocol systematically shapes the jumps and directly connecting to measurable quantities.


\widetext
\pagebreak
\begin{center}
\textbf{\large Supplemental Material: Deterministic Equations for Feedback Control of Open Quantum Systems}
\end{center}
\begin{center}
Alberto J. B. Rosal$^{1,2,*}$, Patrick P. Potts$^3$, and Gabriel T. Landi$^{1,2}$
\end{center}
\begin{center}
$^1$\textit{Department of Physics and Astronomy, University of Rochester, Rochester, New York 14627, USA}\\
$^2$\textit{University of Rochester Center for Coherence and Quantum Science, Rochester, New York 14627, USA}\\
$^3$\textit{Department of Physics and Swiss Nano Science Institute, University of Basel, Klingelbergstrasse 82, 4056 Basel, Switzerland}
\flushleft
\hspace{1.9cm}$^*$ abezerra@ur.rochester.edu
\end{center}

\setcounter{equation}{0}
\setcounter{figure}{0}
\setcounter{table}{0}
\setcounter{page}{1}
\makeatletter
\renewcommand{\theequation}{A\arabic{equation}}
\renewcommand{\thefigure}{A\arabic{figure}}
\renewcommand{\thetable}{A\arabic{table}}

Section~\textcolor{blue}{S.I} introduces the main concepts underlying the main text, including the relation between feedback and data processing, instruments, and quantum-jump detections. We begin with a simple feedback example that illustrates the basic ideas of data processing, followed by additional examples of memory functions. We then show how instruments can describe both dynamical evolution and measurement processes, and briefly discuss how to construct the instruments associated with quantum-jump detections starting from a master equation.
Next, we provide a detailed proof of the central results presented in the main text -- namely, Results 1 and 2 -- in Section~\textcolor{blue}{S.II}. Furthermore, we present a formulation in terms of matrix equations for the main results in Section~\textcolor{blue}{S.III}, where we also include a discussion about the steady state and generator of the feedback dynamics. In Section~\textcolor{blue}{S.IV}, we present the details of the examples used in the main text, and finally in Section~\textcolor{blue}{S.V} we show how some previous results~\cite{PatrickQFPME,Wiseman1994_Feedback,WisemanMilburn1993_Homodyne,FPTLandi} can be recovered starting from Eq.~\eqref{eq: result 1}.

\begin{equation}
\label{eq: result 1}
    \boxed{\varrho_{n+1}(y) = \sum_{x',y'} \delta_{y, f_{n+1}(x',y')} \mathcal{M}_{x'}(y') \varrho_n(y') }
\end{equation}

\begin{table}[h]
\begin{tabular}{cccc}
\hline\hline
Protocol  & Memory                                                         & Instrument                                                    & Equation                                       \\ \hline
   Jump-based time-dependent feedback           & $k_n = x_n  + k_{n-1}\delta_{x_n,0}$
 & $\mathcal{M}_x \rho = V_x \rho V_x^\dagger$                             & Result 2                                                      \\
                       & $\tau_n = \delta_{x_n,0}(\tau_{n-1}+\delta t)$                               &                            &                                                     \\
Quantum Fokker-Planck Equation~\cite{PatrickQFPME} & $y_n = \sum_{j=1}^n \gamma \delta t e^{-\gamma(n-j) \delta t} x_j$           & $\mathcal{M}_z\rho = K_z[ e^{\delta t\mathcal{L}}(\rho)]K_z^\dagger$                 & Eq.~\eqref{eq: FQME}    \\
Single Jump feedback~\cite{Wiseman1994_Feedback}                  & $y_n = x_n$                                                      & $\mathcal{M}_x \rho = \mathcal{F}(x)[V_x(\rho)V_x^\dagger]$              & Eq.~\eqref{eq: single jump feedback equation} \\
Diffusion feedback~\cite{WisemanMilburn1993_Homodyne}                     & $y_n = x_n$                                                      & $\mathcal{M}_z\rho =  \mathcal{F}(z) \{  K_z[ e^{\delta t\mathcal{L}}(\rho)]K_z^\dagger\}$ & Eq.~\eqref{eq: FB + diffusion}\\
Charge-based feedback~\cite{FPTLandi}& $y_n = \sum_{i=1}^n x_i $ & $ \mathcal{M}_x\rho = V_x \rho V_x^\dagger$ & Eq.~\eqref{eq: FPT feedback equation}\\ \hline   \hline          
\end{tabular}
\caption{ This table shows how previous results \cite{PatrickQFPME,Wiseman1994_Feedback,WisemanMilburn1993_Homodyne,FPTLandi} can be recovered starting from Eq.~\eqref{eq: result 1}. Here, the operators $V_x$ describe quantum jump detection, and $ K_z$ represents a Gaussian measurement defined in Eq.~\eqref{eq: kraus operators of gaussian measurements}. The super-operator $\mathcal{F}(x)$, introduced below, is the well-known \textit{feedback super-operator}, and $e^{\delta t \mathcal{L}}$ denotes unitary evolution with $\mathcal{L}\rho = -i[H,\rho]$ for some Hamiltonian $H$.
}
\label{table: summary of results}
\end{table}

\tableofcontents

\section{S.I. Basic Definitions and Preliminaries}
\label{ap:int}

\subsection{A. Data processing, Memory and Feedback}
\label{ap:int_to_signals}
\subsubsection{1. A simple example of feedback}

The purpose of this first section is to introduce readers who may be less familiar with the concept of feedback and data processing. 
As a simple example, let us consider a projective measurement of $\sigma_z$ on a two-level system (qubit). 
This measurement is described by the projectors $\ket{0}\bra{0}$ and $\ket{1}\bra{1}$, where $\sigma_z \ket{l} = (-1)^l \ket{l}$ with $l=0,1$. 
Applying this measurement sequentially at times $t_1,t_2,\cdots,t_n$ produces the dataset $x_{1:n} = (x_1,\cdots,x_n)$, where each outcome $x_i$ is $\pm1$.
At each step, one may apply a unitary operator based on previous measurement outcomes. 
For instance, after $n$ detections, one may consider the unitary
\begin{equation}
    U(x_1,\cdots,x_n) = e^{i g(x_1,\cdots,x_n) \sigma_x}~,
\end{equation}
where $g(x_1,\cdots,x_n)$ is a function of the data, and $\sigma_x$ is the $x$--component of the Pauli matrices. 
Here the measurement provides information about the system, which is then used in the feedback action, namely the unitary transformation. 
The role of the data depends on the task: one may use the full dataset through $g(x_1,\cdots,x_n)$, or a compressed version of it, introducing the notion of \textit{data processing} in feedback dynamics.

To illustrate, suppose the goal is to keep the system in the ground state. 
A simple protocol is: if the average $(1/n)\sum_{i=1}^n x_i$ is positive -- that for a spin--$1/2$ particle means the spin is mostly up (excited state) -- apply a rotation to move it down. 
In this case, one may define
\begin{equation}
    g(x_1,\cdots,x_n) = \alpha~ \theta\left(\frac{1}{n}\sum_{i=1}^n x_i\right)~,
\end{equation}
where $\theta$ is the Heaviside function and $\alpha \in \mathrm{R}$ is a real number. 
Thus, $g(x_1,\cdots,x_n) = \alpha$ if $\frac{1}{n}\sum_{i=1}^n x_i >0$, leading to the unitary $e^{i\alpha\sigma_x}$; otherwise, $g=0$, meaning no action is taken.

We introduce the concept of a \emph{memory function} (also called \emph{stochastic memory}, or simply \emph{memory}) $y_n$, defined as a general function of the previous outcomes, $y_n = y_n(x_{1:n})$. This memory may represent the full dataset, e.g., $y_n = \text{append}(y_{n-1},x_n) = (x_1,\cdots, x_n)$, allowing feedback to depend on all past outcomes. Alternatively, it may encode a compressed version of the data, such as the average $y_n = (1/n)\sum_{i=1}^n x_i$. The choice of memory thus depends on the specific feedback task.
A memory that evolves according to the update rule $y_n = f_n(x_n,y_{n-1})$ is called a \textit{causal memory}, and the function $f_n$ represents the \emph{data-processing function}. It represents how the memory is updated from $y_{n-1}$ to $y_n$ given a new outcome $x_n$. 
In what follows, we present further examples, focusing on two broad classes: linear and non-linear causal memories.

\subsubsection{ 2. Linear memories }
\label{ap: linear signals}
Let us consider that $x_i$ is the outcome of the detection performed at time $t_i$. After $n$ measurements, we have the dataset $x_{1:n} \equiv (x_1,\cdots,x_n)$, and a \textit{linear memory} is defined as a linear combination of the data $x_{1:n}$,
\begin{equation}
\label{eq: linear signals}
    y_n^{\text{lin}} \equiv \sum_{i=1}^n g_{ni} x_i~,
\end{equation}
for a set of coefficients $\{g_{ij}\}$. We can rewrite this memory using matrix notation as follows: one defines $\vec{y}$ as a column vector with components $y_n^{\text{lin}}$, and $\mathbb{G}$ is a matrix with entries $g_{ij}$ such that $g_{ij}=0$ for any $j\geq i+1$. Then, by considering that $\vec{x}$ is a vector with components $x_i$, Eq.~\eqref{eq: linear signals} can be rewritten as
\begin{equation}
    \vec{y} = \mathbb{G}\cdot\vec{x}~. 
\end{equation}

The simplest example is $y_n = x_n$, where the memory corresponds to the current detection, representing a linear memory with $g_{ni} = 0$ for $i \neq n$ and $g_{nn} = 1$.
Another case is the statistical average of the outcomes, defining the \textit{average memory} as
\begin{equation}
\label{eq: average signal}
    y_n^{\text{ave}} = \frac{1}{n}\sum_{i=1}^n x_i.
\end{equation}
Note that this memory satisfies the relation
\begin{equation}
    y_{n}^{\text{ave}} = \frac{1}{n}\left(x_{n}+(n-1)~y_{n-1}^{\text{ave}} \right)~,
\end{equation}
then in this case we have $y_n = f_n(x_n,y_{n-1})$ such that $f_n(x,y) = \left(x+(n-1)~y\right)/n$.

Furthermore, the \emph{low-pass memory} defined as 
\begin{equation}
    y_n^{\text{LP}} = \sum_{i=1}^n \delta t\, \gamma\, e^{-\gamma \delta t(n-i)} x_i
\end{equation}
can be rewritten as
\begin{equation}
   y_n^{\text{LP}} = \sum_{i=1}^n \alpha\, p^{n-i} x_i~, 
\end{equation}
where $\alpha = \gamma \delta t$ and $p = e^{-\alpha}$, and it satisfies
\begin{equation}
    y_n^{\text{LP}} = \alpha x_n + p\, y_{n-1}^{\text{LP}}~.
\end{equation}
Therefore, $y_n^{\text{LP}}$ is not only a causal linear memory, but also time-translation invariant, since the coefficients satisfy $g_{ni} = \alpha\, p^{n-i}$.
Finally, the charge memory, defined as $Q_n = \sum_{i=1}^n x_i$, is an example of a linear memory with uniform weights $g_{ij} = 1$.

\subsubsection{3. Non-linear memories}
A non-linear memory is any function $y_n(x_1,\cdots,x_n)$ that cannot be written as a linear combination of the data $x_i$ as given in Eq.~\eqref{eq: linear signals}. In the main text, we used a non-linear memory to implement a feedback protocol in the context of quantum jump detections. 
In this case, we have a set $\Sigma$ of possible jumps, then the outcomes of a detection at time $t_n = n~\delta t$ is $x_n = 0$ (corresponding to no-jump) or $x_n = k \in \Sigma$ (representing a jump in the channel $k$). The memory that store the last detected jump is given by
\begin{equation}
\label{eq: jump signal}
    k_n = x_n + \delta_{x_n,0}k_{n-1}~,
\end{equation}
where $\delta_{a,b}$ is the Kronecker delta. 

The idea behind Eq.~\eqref{eq: jump signal} is that, in the context of quantum jump detections, the probability of observing a jump is proportional to the infinitesimal interval $\delta t > 0$. As a consequence, jump events are rare, and no-jump evolutions ($x_n = 0$) occur much more frequently than jump events ($x_n = k$). Therefore, a typical detection record takes the form $x_{1:n} = (0,0,0,k,0,0,0,0,k',0,0,0,\ldots)$.
Hence, to discard the trivial no-jump data, the stochastic memory $k_n$ is defined to store the last detected jump as follows: starting from an initial condition $k_0 \in \Sigma$, if a jump is detected at time $t_1$, i.e., $x_1 = k \in \Sigma$, then we update $k_1 = k$; otherwise, we keep $k_1 = k_0$. Equation~\eqref{eq: jump signal} extends this behavior to all $n \geq 1$, ensuring that $k_n$ always stores the last detected jump.

As a second example of non-linear memory, a counting memory that tracks the time elapsed since the last detected jump is defined as
\begin{equation}
\label{eq: counting signal}
    \tau_n = \delta_{x_n,0}(\tau_{n-1} + \delta t)~.
\end{equation}
The idea is that if a no-jump event is detected at time step $n$ (i.e., $x_n = 0$), then the counter is updated as $\tau_n = \tau_{n-1} + \delta t$ for some increment $\delta t > 0$. Otherwise, if a jump occurs, the memory is reset. This allows $\tau_n$ to keep track of the time since the last quantum jump. Similar counting memories can be employed to implement time-dependent feedback strategies that rely on the elapsed time since a given detection event.
Both memories defined in Eqs.~\eqref{eq: jump signal} and \eqref{eq: counting signal} are examples of non-linear causal memories, satisfying the recursive form $y_n = f_n(x_n, y_{n-1})$. For the jump memory~\eqref{eq: jump signal}, the update function is given by $f_n(x, k) = x + \delta_{x,0} \, k$, while for the counting memory~\eqref{eq: counting signal}, it is $f_n(x, \tau) = \delta_{x,0} (\tau + \delta t)$.

\subsection{B. Instruments and Quantum Measurements}
\label{app:00000000A}

Let us consider a collection $\{\mathcal{M}_x\}_{x\in\Sigma}$ of trace non-increasing maps -- $\mathrm{Tr}[\mathcal{M}_x\rho]\leq\mathrm{Tr}[\rho]$ for any density operator $\rho$ -- where $\Sigma$ is an arbitrary set. If the sum $\sum_{x \in \Sigma}\mathcal{M}_x$ is a completely positive trace-preserving (CPTP) map, then the maps $\mathcal{M}_x$ are called \textit{instruments}. Instruments can describe both transitions induced by measurement processes and those arising from dynamical evolutions. First, let us show that any measurement can be written in terms of instruments, and subsequently consider a dynamical evolution combined with detection.

Given a observable $A$ that can assume any value $a$ in the set $\Sigma_{A}$, a measurement of $A$ is defined by a set $\{V_a\}_{a\in\Sigma_A}$ such that 
\begin{equation}
    \sum_{a\in\Sigma_A}V_a^\dagger V_a = \mathbb{1},
\end{equation}
where the set $\{V_a\}_{a\in\Sigma}$ corresponds to the Kraus operators associated with the measurement of $A$, and $\mathbb{1}$ is the identity operator. 
Also, the probability of an outcome $a$ given the state $\rho$ is 
\begin{equation}
    p(a|\rho) = \mathrm{Tr}[V_a\rho V_a^\dagger]~,
\end{equation}
and the post-measurement state given the detenction of outcome $a$ is
\begin{equation}
    \rho_a = \frac{V_a\rho V_a^\dagger}{ \mathrm{Tr}[V_a\rho V_a^\dagger]}~.
\end{equation}
For instance, if one considers a projective measurement of $A$, then $V_a = \left|\psi_a\right>\left<\psi_a\right|$, where $\left|\psi_a\right>$ is the eigenvector of $A$ with eigenvalue $a$.

Now, we can define $\widetilde{\mathcal{M}} \equiv  V_a \rho V_a^\dagger$ for any $a \in \Sigma_A$, hence the probability distribution $p(a|\rho)$ and the post-measurement state $\rho_a$ can be written as
\begin{eqnarray}
    p(a|\rho) &=& \mathrm{Tr}[\widetilde{\mathcal{M}}_a\rho]~,\\
    \rho_a &=& \frac{\widetilde{\mathcal{M}}_a\rho}{ \mathrm{Tr}[\widetilde{\mathcal{M}}_a\rho]}~.
\end{eqnarray}
Let us remember that a set  $\{V_k\}_k$ satisfying the completeness relation $\sum_k V_k^\dagger V_k = \mathbb{1}$ defines a quantum channel (CPTP map) $\Lambda$ by
\begin{equation}
\label{eq: kraus decomposition}
    \Lambda(\rho) = \sum_k V_k \rho V_k^\dagger~,
\end{equation}
where Eq.~\eqref{eq: kraus decomposition} is called \textit{the Kraus decomposition of the channel} $\Lambda$. Thus, we can conclude that $\sum_{a \in \Sigma}\widetilde{\mathcal{M}}_a$ is a CPTP map. Furthermore, since $\sum_{a\in\Sigma_A} V_a^\dagger V_a = \mathbb{1}$ it follows that $ V_a^\dagger V_a \leq \mathbb{1}$, hence $\mathrm{Tr}[\widetilde{\mathcal{M}}_a\rho] \leq \mathrm{Tr}[\rho]$ for any quantum state $\rho$. Therefore, $\widetilde{\mathcal{M}}_a$ is an instrument that describes the detection of the outcome $a \in \Sigma_A$. We can generalize the measurement postulates using the concept of instruments, where a \textit{quantum measurement} is described by a collection $\{\mathcal{M}_x\}_{x \in \Sigma}$ of trace non-increasing super-operators $\mathcal{M}_x$ such that the sum $\sum_{x \in \Sigma}\mathcal{M}_x$ corresponds to a CPTP map, where $\Sigma$ is the set of possible outcomes $x$, the probability distribution of $x$ is given by $p_x = \mathrm{Tr}[\mathcal{M}_x \rho]$, and the system state updates to $\rho \rightarrow \frac{\mathcal{M}_x \rho}{p_x}$ after a measurement of the result $x$.

With the language of instruments, we can also address the dynamical evolution between two detections. For example, let us consider a sequential measurement, where a quantum measurement described by a set $\{V_k\}_k$ of Kraus operators is applied at any time $t_j = j\Delta t$, and let us suppose that the system evolves dynamically from $t_j$ to $t_{j+1}$ according to the a master equation $\partial_t\rho_t = \mathcal{L}\rho_t$, where $\mathcal{L}$ is (for simplicity) time independent. In this case, the instrument describing the joint measurement-dynamics evolution between two detection is given by
\begin{equation}
    \mathcal{M}_x = \widetilde{\mathcal{M}}_xe^{\Delta t\mathcal{L}}~,
\end{equation}
where $\widetilde{\mathcal{M}}_x\rho = V_x \rho V_x^\dagger$ describes the measurement, and $e^{\Delta t\mathcal{L}}$ describes the dynamical evolution of the system. Therefore, when we are dealing with sequential measurements, the instruments $\{\mathcal{M}_x\}$ are also including the possibility of dynamical evolution between two detections. 

\subsection{C. Quantum jump detections and instruments}
\label{ap:quantum_jump_and_inst}
Before presenting the proofs of the main results, it is useful to recall the definitions and notions related to \emph{quantum jump detections}, which form the basis of Result~2 in the main text. To that end, let us consider a system evolving according to a Lindblad master equation
\begin{equation}
\label{QME}
    \frac{d\rho_t}{dt} = \mathcal{L} \rho_t = -i[H,\rho_t]+ ~\sum_{k\in \Sigma} \mathcal{D}[L_k]\rho_t~,
\end{equation}
where $\mathcal{D}[L] \rho \equiv L\rho L^\dagger - \frac{1}{2} \{L^\dagger L, \rho\}$ is the Lindblad dissipator, and $\Sigma$ is a finite alphabet. Here, $L_k$ denote the \textit{jump operators} and $H$ is the system Hamiltonian. Eq.~\eqref{QME} describes the continuous-time evolution of the system state $\rho_t$ in Hilbert space.

On the other hand, Eq.~\eqref{QME} can be rewritten as a discrete infinitesimal evolution from $\rho_t$ to $\rho_{t+\delta t}$
\begin{equation}
\label{eq: unraveling}
    \rho_{t+\delta t} = V_0 \rho_t V_0^\dagger + \sum_{k\in\Sigma} V_k \rho_t V_k^\dagger + \mathcal{O}(\delta t^2)~,
\end{equation}
where only terms up to first order in $\delta t>0$ are retained. Here, $V_0 \equiv \mathbb{1} - i\delta t\, H_{\text{eff}}$, with the effective non-Hermitian Hamiltonian $ H_{\text{eff}} \equiv H - \frac{i}{2}\sum_{k\in\Sigma} L_k^\dagger L_k$, and $V_k \equiv \sqrt{\delta t}\, L_k$ for $k\in\Sigma$. One can verify that
\begin{equation}
    V_0^\dagger V_0 + \sum_{k\in\Sigma} V_k^\dagger V_k = \mathbb{1} + \mathcal{O}(\delta t^2)~.
\end{equation}
Thus, for infinitesimal $\delta t$, the set $\{V_k\}$ constitutes a valid set of Kraus operators. The decomposition of the master equation~\eqref{QME} into this infinitesimal form [Eq.~\eqref{eq: unraveling}] is known as the \emph{unraveling of the master equation}, and defines the \emph{quantum-jump Kraus operators} $V_k$.

Depending on the system, the set $\{V_k\}$ obtained from the unraveling of Eq.~\eqref{QME} may correspond to a physically accessible measurement scheme~\cite{Tutorial,PhysRevLett.57.1699,PhysRevLett.57.1696,PhysRevLett.56.2797,PhysRevLett.54.1023}. In this case, the measurement instruments are given by $\mathcal{M}_k\rho \equiv V_k\rho V_k^\dagger = \delta t\, L_k \rho L_k^\dagger$ for jump detections and $\mathcal{M}_0\rho \equiv (1+\delta t~\mathcal{L}_0)\rho = V_0\rho V_0^\dagger$ for no-jump detections. These measurement schemes are often realized through photon detection, where each jump corresponds to a ``click'' in the detector. The system can either undergo a jump in channel $k$, or remain in the no-jump trajectory and evolve according to the no-jump Liouvillian $\mathcal{L}_0$. The probability of detecting a jump $k$ when the system is in state $\rho$ is 
\begin{equation}
    P(k|\rho) = \mathrm{Tr}[\mathcal{M}_k\rho] = \delta t\, \mathrm{Tr}[L_k\rho L_k^\dagger] \propto \delta t,
\end{equation}
showing that jump events are rare, while no-jump evolutions are much more likely to occur.

\section{S.II. Proof of the main results}
\label{app:00000000B}

\subsection{A. Proof of the Result 1}

Suppose a sequential quantum detection described by the instruments $\{\mathcal{M}_x(y)\}_{x\in \Sigma}$, as defined in the main text, where $y$ denotes a stochastic memory. Let us consider a causal memory governed by the recursive relation
\begin{equation}
    y_{n+1} = f_{n+1}(x_{n+1}, y_n),
\end{equation}
where $f_{n+1}(x,y)$ is a given function. This memory is causal in the sense that it depends on the previous memory $y_n$ and the measurement outcome $x_{n+1}$ at time $t_{n+1} = (n+1)\delta t$. After $n$ measurements, we have the data set $x_{1:n}\equiv(x_1,\cdots,x_n)$, and the conditional state is denoted by $\rho_n(x_{1:n})$. The memory-resolved state is defined as 
\begin{equation}
    \varrho_n(y) \equiv E_{1:n}[\rho_n(x_1,\dots,x_n) \delta_{y,y_n}]~,
\end{equation}
where $E_{1:n}$ represents the ensemble average over all possible trajectories with $n$ outcomes in the data set, and $\delta_{a,b}$ is the Kronecker delta. By using the conditional update rule for the post-measurement state, $\rho_{n+1} = \frac{\mathcal{M}_{x_{n+1}}(y_n)\rho_n(x_{1:n})}{P(x_{n+1}|x_{1:n})}$, we obtain
\begin{align*}
\varrho_{n+1}(y) &= E_{1:n+1}[ \rho_{n+1}(x_{1:n+1}) \delta_{y , y_{n+1}}  ]
\\[0.5cm]
&= E_{1:n+1}\left[\frac{\mathcal{M}_{x_{n+1}}(y_n)\rho_n(x_{1:n})}{P(x_{n+1}|x_{1:n})} \delta_{y , f_{n+1}(x_{n+1},y_n)}\right]~.
\end{align*} 

Let us denote $E_{n+1|1:n}[\cdot]$ as the conditional average over $x_{n+1}$ given the previous outcomes $x_{1:n}$. Using the identity $E_{1:n+1}[\cdot] = E_{1:n}[E_{n+1|1:n}[\cdot]]$, we find
\begin{align*}
\varrho_{n+1}(y) &=E_{1:n}\left\{ \sum_{x'} \delta_{y ,f_{n+1}(x',y_n)}\mathcal{M}_{x'}(y_n)\rho_n(x_{1:n}) \right\}
\\[0.5cm]
&=\sum_{x'}E_{1:n}\left[ \delta_{y , f_{n+1}(x',y_n)}\mathcal{M}_{x'}(y_n)\rho_n(x_{1:n}) \right]~.
\end{align*}
To isolate the definition of $\varrho_n(y)$, we introduce a dummy summation over $y'$ and a Kronecker delta as follows
\begin{align*}
\varrho_{n+1}(y) &= \sum_{x'} \sum_{y'} E_{1:n}\left\{ \delta_{y , f_{n+1}(x',y')} \delta_{y' , y_n}\mathcal{M}_{x'}(y') \rho_n(x_{1:n}) \right\}~,
\end{align*}
where we have inserted a Kronecker delta along with a sum over $y'$, then replaced $y_n$ in the expression with $y'$ via $\delta_{y', y_n}$. Notably, this also allows us to replace $y_n$ in both $\mathcal{M}_{x'}(y_n)$ and $f_{n+1}(x',y_n)$ with the deterministic value $y'$, yielding
\begin{equation}
    \varrho_{n+1}(y) = \sum_{x',y'}\delta_{y , f_{n+1}(x',y')}\mathcal{M}_{x'}(y')E_{1:n}\left\{ \rho_n(x_{1:n}) \delta_{y' , y_n} \right\}~.
\end{equation}
Finally, we identify the definition of the memory-resolved state as $\varrho_n(y') = E_{1:n}\left\{ \rho_n(x_{1:n}) \delta_{y' , y_n} \right\}$, and we arrive at
\begin{equation}
    \varrho_{n+1}(y) = \sum_{x',y'} \delta_{y , f_{n+1}(x',y')} \mathcal{M}_{x'}(y') \varrho_n(y')~,
\end{equation}
as described in Result~1.

For simplicity, we have so far assumed that both the memory $y_n$ and the outcomes $x_n$ take values in discrete sets. The case of continuous variables can be treated analogously.  
For a continuous-valued memory, we define the memory-resolved state as
\begin{equation}
    \varrho_n(y) \equiv E_{1:n}\!\left[\rho_n(x_1,\dots,x_n) \, \delta(y-y_n)\right]~,
\end{equation}
where $\delta(y-y_n)$ denotes the Dirac delta.  
In this case, the sums are replaced by integrals, and Result~1 for continuous outcomes and a continuous memory becomes
\begin{equation}
    \varrho_{n+1}(y) = \int dx' \, dy' \, \delta\!\left(y - f_{n+1}(x',y')\right)\mathcal{M}_{x'}(y') \varrho_n(y')~,
\end{equation}
where $\varrho_n(y)$ is now the probability density function of the continuous-valued memory $y_n$ at time $t_n = n\delta t$.

\subsection{B. Proof of the Result 2}
\subsubsection{1. Jump-based time-dependent feedback}
Result~(1) describes the general evolution of a feedback dynamics defined by the instruments $\mathcal{M}_x(y)$ and a causal memory $y_n$. Importantly, this result also holds when more than one memory is considered, i.e. when $y_n^{(i)} = f_n^{(i)}(x_n,y_{n-1}^{(i)})$ with $i=1,2,\dots$. 
In this case, the total memory $y_n = (y_n^{(1)}, y_n^{(2)}, \cdots) $ can be regarded as a vector of memory functions $y_n^{(i)}$.
For instance, if two memories ($i=1,2$) are included in the feedback strategy, the deterministic equation becomes
\begin{equation}
\label{eq: deterministic feedback with two signals}
\varrho_{n+1}(y_a,y_b) = \sum_{x',y_a',y_b'}
\delta_{y_a, f_{n+1}^{(1)}(x',y_a')}
\delta_{y_b, f_{n+1}^{(2)}(x',y_b')}
\mathcal{M}_{x'}(y_a',y_b'), \varrho_n(y_a',y_b')~,
\end{equation}
where $y_a$ and $y_b$ denote realizations of the memories $y_n^{(1)}$ and $y_n^{(2)}$, respectively. 
In general, for each additional memory $y_n^{(i)}$ included in the feedback dynamics, the sum acquires an extra Kronecker delta enforcing the corresponding update function $f_n^{(i)}$.


Suppose that we are applying a sequential measurement of quantum jumps described by the Kraus operators $V_k = \sqrt{\delta t}L_k$ for $k \in \Sigma$, and $V_0 = \mathbb{1} - i \delta t H_{\text{eff}}$, where $H_{\text{eff}} = H - \frac{i}{2}\sum_{k\in \Sigma} L_k^\dagger L_k$ is the effective Hamiltonian, and $\Sigma$ is the set of all possible jumps. Hence, the correspondent instrument is defined by $\mathcal{M}_x \rho = V_x \rho V_x^\dagger$. Consider a feedback scheme conditioned on the last detected quantum jump \textit{and} the time interval since its occurrence. The stochastic memory that records the last jump is given by 
\begin{equation}
    \label{eq: jump filter appendix}
    k_n = x_n  + k_{n-1}\delta_{x_n,0},
\end{equation}
where $\delta_{x_n,0}$ is the Kronecker delta, and $x_n$ represents the outcome of the quantum jump detection at time $t_n =  n \delta t $. In this context, $x_n = 0$ corresponds to a no-jump detection, while $x_n = k$ indicates that a quantum jump occurred in channel $k$. Furthermore, the memory that tracks the time elapsed since the last detected jump is given by
\begin{equation}
    \label{eq: counting filter appendix}
    \tau_n = \delta_{x_n,0}(\tau_{n-1}+\delta t).
\end{equation}
We are considering generic initial condition for both memories, where $k_0 = \bar{k} \in \Sigma$ is fixed and $\tau_0 = 0$. It is equivalent to say that the system is prepared such that we have a jump $\bar{k}$ at time $t = 0$. Hence, the feedback action is implemented in the quantum jump detections at time $t_n = n\delta t$ by considering the Kraus operators $V_{k}(k_{n-1},\tau_{n-1})$.

Concretely, both the jump memory in Eq.~\eqref{eq: jump filter appendix} and the counting memory in Eq.~\eqref{eq: counting filter appendix} satisfy the update rules $k_n = f_1(x_n,k_{n-1})$ and $\tau_n = f_2(x_n,\tau_{n-1})$, with $f_1(x,k) = x + k \delta_{x,0}$ and $f_2(x,\tau) = \delta_{x,0}(\tau+\delta t)$.
In this case, the full memory is simply the pair $y_n = (k_n,\tau_n)$, and applying the general prescription of Eq.~\eqref{eq: deterministic feedback with two signals} to these two memories, we obtain
 \begin{eqnarray}
        \varrho_{n+1}(k,\tau) &=& \sum_{x',k',\tau'} \delta_{k,x'+k'\delta_{x',0}} \delta_{\tau,\delta_{x',0}(\tau'+\delta t)} \mathcal{M}_{x'}(k',\tau') \varrho_n(k',\tau')\\
         &=& \sum_{k',\tau'} \delta_{k,k'} \delta_{\tau,(\tau'+\delta t)} \mathcal{M}_0(k',\tau')\varrho_n(k',\tau') + \sum_{x' \in \Sigma}\sum_{k',\tau'}  \delta_{k,x'}\delta_{\tau,0} \mathcal{M}_{x'}(k',\tau') \varrho_n(k',\tau')\\
         &=&\label{eq: discrete result 2} \sum_{h=0}^n \delta_{\tau,(h\delta t+\delta t)} \mathcal{M}_0(k,h\delta t)\varrho_n(k,h\delta t) + \delta_{\tau,0} \sum_{k'\in\Sigma}\sum_{h = 0}^n   \mathcal{M}_{k}(k',h\delta t) \varrho_n(k',h\delta t)~.
    \end{eqnarray}
The first equality follows directly from Result 1 [Eq.~\eqref{eq: result 1}]. In the second equality, we decompose the sum over $x'$ into the two possible cases: $x' = 0$ (no-jump detection) and $x' \in \Sigma$ (jump detection). The transition from the second to the third line is obtained by using the Kronecker delta to evaluate the first sum over $k'$ and the last sum over $x'$. Finally, in Eq.~\eqref{eq: discrete result 2}, we make the summation ranges explicit: $k' \in \Sigma$ and $\tau' = h  \delta t$ with $h = 0, 1, \dots, n$.

Note that Eq.~\eqref{eq: discrete result 2} can be separated into two cases: $\tau = 0$ and $\tau > 0$. When $\tau = 0$, only the second term survives, yielding
\begin{equation}
\label{eq: case tau = 0}
    \varrho_{n+1}(k,0) = \sum_{k'\in\Sigma}\sum_{h=0}^n \delta t \mathcal{J}_k(k',h\delta t)\varrho_n(k',h\delta t)~,
\end{equation}
where $\mathcal{M}_k \equiv \delta t \mathcal{J}_k$ for $k \in \Sigma$, and $\mathcal{J}_k\varrho = L_k \varrho L_k^\dagger$ represents the channel $k\in\Sigma$. On the other hand, for $\tau>0$, namely $\tau = p\delta t$ where $p = 1,\cdots,n+1$, only the first term of Eq.~\eqref{eq: discrete result 2} survives, and we have
\begin{equation}
    \varrho_{n+1}(k,p\delta t) = \sum_{h=0}^n \delta_{p \delta t,(h\delta t+\delta t)} \mathcal{M}_0(k,h\delta t)\varrho_n(k,h\delta t) = \mathcal{M}_0(k,p\delta t - \delta t) \varrho_n(k, p\delta t - \delta t) = e^{\delta t \mathcal{L}_0(k,(p-1)\delta t)}\varrho_n(k,(p-1)\delta t)~,
\end{equation}
where we used the Kronecker delta to perform the sum over $h$, and $\mathcal{M}_0\varrho = V_0\varrho V_0^\dagger = e^{\delta t \mathcal{L}_0}\varrho$ for quantum jumps. Note that $p = 1,\cdots, n+1$ ensures $(p-1)\delta t \geq 0$. Then, by applying the last equation recursively, we obtain
\begin{equation}
\label{eq: case tau != 0}
    \varrho_{n+1}(k,p\delta t) = e^{\delta t \mathcal{L}_0(k,(p-1)\delta t)} e^{\delta t \mathcal{L}_0(k,(p-2)\delta t)} \cdots e^{\delta t \mathcal{L}_0(k,0)} \varrho_{n+1-p}(k,0)~.
\end{equation}

By taking the continuous measurement limit $\delta t \rightarrow 0$ and $p \rightarrow \infty$ such that $\tau = p\delta t$ remains fixed (and similarly $t = n\,\delta t$), and that $\varrho_n(k, p\delta t) \rightarrow \varrho_t(k, \tau)$, Eq.~\eqref{eq: case tau != 0} becomes
\begin{equation}
\label{eq: result 2 eq1}
    \varrho_t(k,\tau) = \mathcal{T}\left[ e^{\int_0^\tau ds \mathcal{L}_0(k,s)} \right ]\varrho_{t-\tau}(k,0)~.
\end{equation}
Finally, by applying this same limit in Eq.~\eqref{eq: case tau = 0}, we have
\begin{equation}
    \varrho_t(k,0) = 2 \delta(t) \delta_{k,\bar{k}}\bar{\rho}_0+ \sum_{k'\in\Sigma} \int_0^t d\tau' \mathcal{J}_k(k',\tau')\varrho_t(k',\tau')~,
\end{equation}
where $2\delta(t)\delta_{k,\bar{k}}\bar{\rho}_0$ accounts for the initial condition. The convention $\int_0^t d\tau\delta(t-\tau)=1/2$ is used since the Dirac delta lies at the integration boundary.
By using Eq.~\eqref{eq: case tau != 0} we finally have
\begin{equation}
\label{eq: result 2 eq2}
     \varrho_t(k,0) = 2 \delta(t) \delta_{k,\bar{k}}\bar{\rho}_0+ \sum_{k'\in\Sigma} \int_0^t d\tau' \mathcal{J}_k(k',\tau')\mathcal{T}\left[ e^{\int_0^{\tau'} ds \mathcal{L}_0(k',s)} \right ]\varrho_{t-\tau'}(k',0)~.
\end{equation}
Note that Eq.~\eqref{eq: result 2 eq2} represents a sum over all possible previous trajectories: a jump of type $k$ can occur at time $t>0$ if it was preceded by a jump of type $k'$ at time $t - \tau'$, followed by no jumps in the interval $(t - \tau', t]$. The sum over $k'$ and the integral over $\tau'$ account for all such possibilities. Eqs.~\eqref{eq: result 2 eq1} and \eqref{eq: result 2 eq2} describe the feedback dynamics under jump monitoring, thus completing the proof of Result~2.

Next, we analyze the particularly interesting case where feedback depends only on the quantum jumps, ignoring the time elapsed since the last detection in the feedback action.

\subsubsection{2. Jump-based time-independent feedback}
Let us consider feedback that depends solely on the last detected jump, so that the instruments can be written as $\mathcal{M}_x(k,\tau) \rightarrow \mathcal{M}_x(k)$. 
The corresponding memory-resolved state is obtained by marginalizing over the time variable: 
$\varrho_t(k) = \int_0^t d\tau \, \varrho_t(k,\tau)$.
Therefore, by applying this marginalization on both sides of Eq.~\eqref{eq: discrete result 2}, we have
\begin{eqnarray}
    \varrho_{n+1}(k) &=& \mathcal{M}_0(k) \varrho_n(k) +\sum_{k'\in\Sigma}  \mathcal{M}_{k}(k') \varrho_n(k')\\
    &=& V_0(k)\varrho_n(k)V_0^\dagger(k) + \delta t \sum_{k'\in\Sigma}L_k(k')\varrho_n(k')L_k^\dagger(k')~,
\end{eqnarray}
where in the second equality we have used the instruments related with quantum jump detections. For the first term, we have
\begin{equation}
     V_0(k)\varrho_n(k)V_0^\dagger(k) = (\mathbb{1}+\delta t\mathcal{L}_0(k))\varrho_n(k) = \varrho_n(k) -i\delta t [H(k),\varrho_n(k)] -\frac{\delta t}{2}\sum_{k'\in\Sigma}\{L_{k'}^\dagger(k)L_{k'}(k),\varrho_n(k)\}~,
\end{equation}
then we have
\begin{equation}
    \varrho_{n+1}(k) = \varrho_n(k)  -i\delta t [H(k),\varrho_n(k)] -\frac{\delta t}{2}\sum_{k'\in\Sigma}\{L_{k'}^\dagger(k)L_{k'}(k),\varrho_n(k)\} +  \delta t \sum_{k'\in\Sigma}L_k(k')\varrho_n(k')L_k^\dagger(k')~.
\end{equation}

By taking the continuous measurement limit $\delta t \rightarrow 0$, where $\varrho_n(k)\rightarrow\varrho_t(k)$ and $(\varrho_{n+1}(k) - \varrho_{n}(k))/\delta t \rightarrow \partial_t \varrho_t(k)$, we finally obtain 
\begin{eqnarray}
\label{eq: Patrick's equation appendix}
    \partial_t \varrho_t(k) = -i \big[H(k),\varrho_t(k)\big]  +\sum_{k' \in \Sigma}\left(\ L_k^{}(k') \varrho_t(k') L_k^\dagger(k') - \frac{1}{2} \{L_{k'}^\dagger(k)L_{k'}(k),\varrho_t(k)\}\right). 
\end{eqnarray}
Therefore, the master equation given by Eq.~\eqref{eq: Patrick's equation appendix} provides the feedback dynamics when only the last jump is considered. We can recover the unconditional state of the system by marginalizing the memory-resolved state over $k$, $\bar{\rho}_t = \sum_{k\in\Sigma}\varrho_t(k)$, and the probability of detecting a jump $k$ at time $t$ is given by $\mathrm{Tr}[\varrho_t(k)]$.

\section{S.III. Matrix equations: generators and steady state of feedback dynamics}
\label{app:00000000C}

We can rewrite our results in terms of matrix equations as follows. Let us define the propagator 
\begin{equation}
    \Lambda_y^{(n)}(y') \equiv \sum_{x'} \delta_{y, f_{n+1}(x', y')} \mathcal{M}_{x'}(y')~,
\end{equation}
so that Eq.~\eqref{eq: result 1} takes the form 
\begin{eqnarray}
    \varrho_{n+1}(y) &=& \sum_{x',y'} \delta_{y, f_{n+1}(x',y')} \mathcal{M}_{x'}(y') \varrho_n(y') = \sum_{y'}\left( \sum_{x'}\delta_{y, f_{n+1}(x',y')} \mathcal{M}_{x'}(y')\right) \varrho_n(y')\\
    \label{eq: result 1 in component form}
    &=& \sum_{y'} \Lambda_y^{(n)}(y') \varrho_{n}(y')~.
\end{eqnarray}
Since the maps $\mathcal{M}_x(y')$ are instruments, then $\sum_y \Lambda_y^{(n)}(y') = \sum_{x'} \mathcal{M}_{x'}(y')$ is a completely positive and trace-preserving (CPTP) map for any $y'$, ensuring consistency with the general evolution of a hybrid quantum state~\cite{PhysRevX.13.041040}. 
This new representation of Eq.~\eqref{eq: result 1} resembles a matrix product. To make this structure explicit, let us define the column vector $\vec{\varrho}_n$, whose components correspond to the memory-resolved states $\varrho_n(y)$, with $y$ ranging over the possible values of the stochastic memory $y_n$ at time $t_n = n \delta t$. We also introduce the double-indexed object $\mathbf{\Lambda_n} \equiv [ \Lambda_y^{(n)}(y')]_{y,y'}$, which represents a matrix whose components are the super-operators $\Lambda_y^{(n)}(y')$. 
With this notation, the right-hand side of Eq.~\eqref{eq: result 1 in component form} can be written as the $y$--component of a matrix product, $\sum_{y'} \Lambda_y^{(n)}(y') \varrho_{n}(y') = (\mathbf{\Lambda}_n\cdot\vec{\varrho}_{n})_y$, where the left-hand side corresponds to the $y$--component of $\vec{\varrho}_{n+1}$. Therefore, using this matrix representation, Eq.~\eqref{eq: result 1 in component form} takes the compact form 
 \begin{equation}
 \label{eq: deterministic eq1 in matrix form}
 \vec{\varrho}_{n+1} = \mathbf{\Lambda_n}\cdot\vec{\varrho}_{n}~.
\end{equation}

Moreover, we can define the generator of the feedback dynamics as follows. Note that $\vec{\varrho}_{n+1} - \vec{\varrho}_{n} = (\mathbf{\Lambda_n} - I_d)\vec{\varrho}_{n} $, where $I_d$ represents the identity. Hence, if $\lim_{\Delta t \rightarrow0} (\mathbf{\Lambda_n} - I_d)/\Delta t$ is well defined, we can introduce the generator $\mathcal{L}_{\text{fb}} = \lim_{\Delta t \rightarrow0} (\mathbf{\Lambda_n} - I_d)/\Delta t $, then $\partial_t \vec{\varrho}_t = \mathcal{L}_{\text{fb}} \cdot \vec{\varrho}_t$. Finally, the memory-resolved steady state is defined as $\vec{\varrho}_{ss} \equiv \lim_{t\rightarrow\infty}\vec{\varrho}_t$, or equivalently $\varrho_{ss}(y) = \lim_{t\rightarrow\infty}\varrho_t(y)$.
By considering Eq.~\eqref{eq: deterministic eq1 in matrix form}, we have
 \begin{equation}
     \label{eq: general steady state in matrix form}
     \vec{\varrho}_{ss} =  \mathbf{\Lambda_\infty}\cdot\vec{\varrho}_{ss}~,
 \end{equation}
 where  $\mathbf{\Lambda_\infty} = \lim_{n\rightarrow\infty}\mathbf{\Lambda_n}$. Hence, one find that the memory-resolved steady state can be seen as the eigenvector of $ \mathbf{\Lambda_\infty}$ with unit eigenvalue.

 One can find the steady state of the system in a jump-based feedback by considering the limit $t \rightarrow\infty$ in Result~2. 
We assume that $\mathcal{L}_0(k,\tau)$ has only eigenvalues with negative real part \cite{Tutorial}, so it can be decomposed as $\mathcal{L}_0(k,\tau) = - \mathrm{R}(k,\tau) + i \mathrm{I}(k,\tau)$, where $\mathrm{R}(k,\tau)$ is a super-operator with positive eigenvalues and $\mathrm{I}(k,\tau)$ corresponds to a super-operator with real eigenvalues. In the vectorization notation \cite{Tutorial}, states are mapped to vectors and super-operators to matrices; thus, $\mathrm{R}(k,\tau)$ corresponds to a positive matrix. Consequently,
\begin{equation}
    G(k',\tau')=\mathcal{T}\!\left[e^{\int_0^{\tau'} ds\,\mathcal{L}_0(k',s)}\right] 
= \mathcal{T}\!\left[e^{- \int_0^{\tau'} ds\,\mathrm{R}(k,s) +i \int_0^{\tau'} ds\,\mathrm{I}(k,s)}\right],
\end{equation}
which features two distinct contributions: an exponential decay from $-\int_0^{\tau'} ds\,\mathrm{R}(k,s)$ and a phase factor from $i \int_0^{\tau'} ds\,\mathrm{I}(k,s)$. In this case, $G(k',\tau')=\mathcal{T}\left[e^{\int_0^{\tau'} ds\mathcal{L}_0(k',s)}\right]$ decays to zero for large $\tau'$. 
Thus, in the limit $t \rightarrow \infty$, Eq.~\eqref{eq: result 2 eq2} yields 
\begin{eqnarray}
     \varrho_{ss}(k,0) &\equiv& \lim_{t\rightarrow\infty}\varrho_t(k,0) 
     = \sum_{k'\in\Sigma} \lim_{t\rightarrow\infty}\int_0^t d\tau'\,
     \mathcal{J}_k(k',\tau')\, G(k',\tau')\, \varrho_{t-\tau'}(k',0)~,
\end{eqnarray}
and since $G(k',\tau')$ vanishes as $\tau' \to \infty$, one can consider large but finite values of $\tau'$ and extend the upper integration limit to infinity, allowing the replacement $\varrho_{t-\tau'}(k',0) \rightarrow \varrho_{ss}(k',0)$ inside the integral.
This gives
\begin{equation}
     \label{eq: steady state for combined signal 1}
      \varrho_{ss}(k,0) 
      = \sum_{k'\in\Sigma}\left(\int_0^\infty d\tau'\, \mathcal{J}_k(k',\tau')\, G(k',\tau')\right) 
      \varrho_{ss}(k',0).
\end{equation}

Following the same way as in Eq.~\eqref{eq: deterministic eq1 in matrix form}, we can rewrite Eq.~\eqref{eq: steady state for combined signal 1} in a matrix form as follows. Let us define the super-operator
\begin{equation}
    \mathbf{\Lambda_\infty} \equiv \left[\int_0^\infty d\tau' \, \mathcal{J}_k(k',\tau')\, G(k',\tau')\right]_{k,k'}~,
\end{equation}
and let $\vec{\varrho}_{ss}(0)$ denote the column vector with components $\varrho_{ss}(k,0)$ for $k \in \Sigma$. 
Hence, Eq.~\eqref{eq: steady state for combined signal 1} can be rewritten as
\begin{equation}
      \varrho_{ss}(k,0) 
      = \sum_{k'\in\Sigma} \left(\Lambda_{\infty}\right)_{k,k'} \varrho_{ss}(k',0)
      = \left(\Lambda_{\infty} \cdot \vec{\varrho}_{ss}(0)\right)_{k}~.
\end{equation}
Therefore, Eq.~\eqref{eq: steady state for combined signal 1} becomes $\vec{\varrho}_{ss}(0) = \mathbf{\Lambda_\infty}\cdot \vec{\varrho}_{ss}(0)$, where the steady state is the eigenvector of $\mathbf{\Lambda_\infty}$ corresponding to the eigenvalue 1, as discussed in Eq.~\eqref{eq: general steady state in matrix form}. Consequently, the eigenvector of $\mathbf{\Lambda}_\infty$ provides us $\varrho_{ss}(k,0)$. On the other hand, Eq.~\eqref{eq: result 2 eq1} gives $\varrho_t(k,\tau)$, where we have
\begin{equation}
    \varrho_{ss}(k,\tau) = \mathcal{T}\left[ e^{\int_0^\tau ds \mathcal{L}_0(k,s)} \right ]\varrho_{ss}(k,0)~.
\end{equation}
Hence, in the continuous monitoring limit $\delta t\rightarrow0$, the unconditional steady state is given by $\bar{\rho}_{ss} = \sum_k \int d\tau~\varrho_{ss}(k, \tau)$.

Now, let us find the steady state for the jump feedback, namely when the feedback is based only on the most recent jump. In this case, the feedback dynamics is described by Eq.~\eqref{eq: Patrick's equation appendix}. Note that we can rewrite this equation as 
\begin{equation}
    \frac{\partial \varrho_t(k)}{\partial t} = \Phi(k) \varrho_t(k) + \sum_{k' \in \Sigma} L_k(k') \varrho_t(k') L_k^\dagger(k')~,
\end{equation}
where
\begin{equation}
    \Phi(k) \varrho_t(k) \equiv -i[H(k), \varrho_t(k)]  - \frac{1}{2} \sum_{k' \in \Sigma} \left\{ L_{k'}^\dagger(k) L_{k'}(k), \varrho_t(k) \right\}.
\end{equation}
The term $\sum_{k' \in \Sigma} L_k(k') \varrho_t(k') L_k^\dagger(k')$ couples different jumps in the differential equation of the memory-resolved state $\varrho_t(k)$, and describes the contribution of various jump channels due to the feedback action.

Let us define a column vector $\vec{\varrho}_t$ whose components are the elements $\varrho_t(k)$ for each jump $k \in \Sigma$. Hence, the previous equation can be compactly expressed as
\begin{equation}
    \frac{\partial \vec{\varrho}_t}{\partial t} = \Psi \vec{\varrho}_t + \Xi \vec{\varrho}_t~,
\end{equation}
where $\Psi$ is a diagonal matrix whose diagonal elements are given by the super-operators $\Phi$(k), i.e., $[\Psi]_{k\, k'} = \Phi(k) \delta_{k\,k'}$ with $k,\,k'\in\Sigma$, and $\Xi$ is a matrix whose entries are the super-operators $\mathcal{J}_k(k')$, i.e., $[\Xi]_{k\, k'} = \mathcal{J}_k(k')$. Defining $\Pi \equiv \Psi + \Xi$, we finally obtain
\begin{equation}
\label{eq: vectorial version of Result 2}
    \frac{\partial \vec{\varrho}_t}{\partial t} = \Pi \vec{\varrho}_t~.
\end{equation}
Therefore, Eq.~\eqref{eq: vectorial version of Result 2} provides a compact, vectorial formulation of Result 2 for the memory-resolved state when the feedback is based solely on the last detected jump. In the steady state, where $\partial_t \vec{\varrho}_t = 0$, the solution satisfies the eigenvalue equation
\begin{equation}
    \Pi \vec{\varrho}_{ss} = 0~,
\end{equation}
providing the memory-resolved steady state $\varrho_t(k)$ for each jump $k\in\Sigma$. As before, the unconditional steady state in this continuous monitoring limit is given by $\bar{\rho}_{ss} = \sum_{k\in\Sigma}\varrho_{ss}(k)$.

\section{S.IV. Examples and details}
\label{app:00000000D}

\subsection{A. Rotating frame and Rotating Wave Approximation}
In the example presented in the main text, we used a rotating frame together with the rotating wave approximation. In this section, we briefly review these concepts. Consider a two-level system with transition frequency $\omega$, driven by an external field with frequency $\omega_d$ and coupling strength $\lambda$. The Hamiltonian in the Schrodinger picture is given by ($\hbar = 1$)
\begin{equation}
    H = -\frac{\omega}{2}\sigma_z + \Omega  \cos{(t\omega_d )}\sigma_x~,
\end{equation}
where $\sigma_{z,x}$ are the Pauli matrices. In this convention, the ground state $\left|g\right>$ has energy $-\omega/2$. The Hamiltonian in the rotating frame is defined as
\begin{equation}
    H' \equiv e^{\frac{-i t \omega_d }{2}\sigma_z} H e^{\frac{i t \omega_d }{2}\sigma_z} + \frac{  \omega_d }{2}\sigma_z = -\frac{\Delta}{2}\sigma_z + \frac{\Omega}{2}(\sigma_+ + \sigma_-) + \frac{\Omega}{2}(e^{-2i\omega_d t}\sigma_+ +e^{2i\omega_dt}\sigma_-  )~,
\end{equation}
where we used that $e^{-it\omega_d\sigma_z/2} \sigma_{\pm}e^{it\omega_d\sigma_z/2} = e^{\mp it\omega_d}\sigma_{\pm}$, $\sigma_+ = \left|e\right>\left<g\right|$, $\sigma_- = \left|g\right>\left<e\right|$, and $\Delta = \omega-\omega_d$.

The rotating wave approximation consists of neglecting the rapidly oscillating terms in the Hamiltonian, which average out over time and have negligible effect on the system's dynamics. Hence, taking $\frac{\Omega}{2}(e^{-2i\omega_dt}\sigma_+ +e^{+2i\omega_dt}\sigma_-  )\approx0$, the Hamiltonian becomes
\begin{equation}
    H' = -\frac{\Delta}{2}\sigma_z + \lambda \sigma_x,
\end{equation}
where $\lambda \equiv \Omega/2$, and $\sigma_x = \sigma_+ + \sigma_-$. To return to the Schrodinger picture, we need to apply the inverse transformation, where 
\begin{equation}
    \label{eq: rotating frame transformation}
    \rho = e^{\frac{i t \omega_d }{2}\sigma_z} \rho'e^{\frac{-i t \omega_d }{2}\sigma_z}.
\end{equation}

\subsection{B. Example 1: inversion protocol}
In the main text, we considered a two-level system of energy $\omega$ ($\hbar = 1$) coupled to a thermal bath under the action of feedback. By employing a time-dependent, jump-based feedback protocol described by Result~2, we demonstrated that it is possible not only to increase the excited state population compared to the thermal case without feedback, but also to achieve population inversion, with $P_e > P_g$. In this section, we provide a step-by-step derivation of the equations governing the populations under feedback. We consider continuous monitoring of quantum jumps, described by the jump operators
\begin{equation}
L_{-} = \sqrt{\gamma(\bar{N}+1)} ~|g\rangle\langle e|, \quad
L_{+} = \sqrt{\gamma\bar{N}} ~|e\rangle\langle g|,
\end{equation}
which correspond to the emission ($x_n = -1$) and absorption ($x_n = +1$) of a quanta, respectively. The outcome $x_n = 0$ represents a no-jump detection.

The feedback action is defined as follows: if an emission is detected, we turn on an external drive with frequency $\omega_d$ for a total duration $\tau_1$, after a time delay $\tau_0$ without absorption. During this feedback window, and working in a rotating frame under the rotating wave approximation, the Hamiltonian becomes
\begin{equation}
    H_{\text{on}} = -\frac{\Delta}{2}\sigma_z + \lambda \sigma_x~, 
\end{equation}
where $\Delta = \omega - \omega_d$, otherwise we remove the drive ($\lambda\rightarrow0$), and we obtain the thermal qubit with Hamiltonian 
\begin{equation}
    H_{\text{off}} = -\frac{\Delta}{2}\sigma_z~.
\end{equation}
Since this feedback protocol affects only the system's Hamiltonian, then the unconditional state evolves according to
\begin{equation}
\label{eq: app unconditional evolution for example}
    \bar{\rho}_t = G(\bar{k},t) \bar{\rho}_0 + \sum_{k\in\Sigma}\int_0^td\tau 
   \mathcal{T}\left[e^{\int_0^{\tau} ds\mathcal{L}_0(k,s)}\right]
    \mathcal{J}_k \bar{\rho}_{t-\tau}~,
\end{equation}
and the steady-state, if it exists, is the solution of the algebraic equation
\begin{equation}
\label{eq: app unconditional steady state for Result 2}
    \bar{\rho}_{\rm ss} = \Omega \bar{\rho}_{\rm ss},
    \qquad 
    \Omega = \sum_{k\in\{+1,-1\}}\int_0^\infty d\tau ~
   \mathcal{T}\left[ e^{\int_0^\tau ds \mathcal{L}_0(k,s)} \right ]
    \mathcal{J}_k.
\end{equation}

In order to verify whether a steady state exists, it is sufficient to show that the no-jump Liouvillian has only eigenvalues with negative real parts. The reasoning is as follows: if $\mathcal{L}_0(k,\tau)$ has only negative eigenvalues, then the term $\mathcal{T}\left[e^{\int_0^{\tau} ds\,\mathcal{L}_0(k,s)}\right]$ vanishes as $\tau \rightarrow \infty$. Therefore, in the steady-state limit of Eq.~\eqref{eq: app unconditional evolution for example}, when $t \rightarrow \infty$, only contributions with finite $\tau$ survive in $\mathcal{T}\left[e^{\int_0^{\tau} ds\,\mathcal{L}_0(k,s)}\right]$, and we can replace $\rho_{t-\tau} \rightarrow \bar{\rho}_{ss}$ inside the integral. For this feedback protocol, the no-jump Liouvillian are given by
\begin{equation}
\label{eq: app no-jump liouvilian of the example}
    \mathcal{L}_0(k,s) = \begin{cases}
    \mathcal{L}_0^{\text{on}}  & \text{if }  k=-1\text{ and }  s \in[\tau_0,\tau_0+\tau_1] ,\\
 \mathcal{L}_0^{\text{off}} & \text{otherwise } , 
    \end{cases}
\end{equation}
where $\mathcal{L}_0^{\text{off}}$ and $\mathcal{L}_0^{\text{on}}$ are the no-jump Liouvillian related with $H_{\text{off}}$ and $H_{\text{on}}$, respectively. By applying vectorization~\cite{Tutorial}, each $2\times2$ density matrix $\rho$ is mapped to a four-component column vector, and the Liouvillian becomes a $4\times4$ matrix acting on these vectors.
For a resonant external drive ($\Delta = 0$), the no-jump Liouvillians are given by
\begin{align}
\mathcal{L}_0^{\text{off}}&= 
\begin{pmatrix}
-\bar{N} \gamma & 0 & 0 & 0 \\
0 & -\frac{1}{2} (1 + 2 \bar{N}) \gamma  & 0 & 0 \\
0 & 0 & -\frac{1}{2} (1 + 2 \bar{N}) \gamma  & 0 \\
0 & 0 & 0 & -(1 + \bar{N}) \gamma
\end{pmatrix}, \\
\mathcal{L}_0^{\text{on}} &= 
\begin{pmatrix}
-\bar{N} \gamma & -i \lambda & i \lambda & 0 \\
-i \lambda & -\frac{1}{2}(1 + 2 \bar{N}) \gamma & 0 & i \lambda \\
i \lambda & 0 & -\frac{1}{2}(1 + 2 \bar{N}) \gamma & -i \lambda \\
0 & i \lambda & -i \lambda & -(1 + \bar{N}) \gamma
\end{pmatrix}~,
\end{align}
and both exhibit eigenvalues with negative real parts. Hence, a steady state exists and is given by Eq.~\eqref{eq: app unconditional steady state for Result 2}.
We can compute $\Omega$ directly from the definition by using Eq.~\eqref{eq: app no-jump liouvilian of the example}, and one finds 
\begin{equation}
    \Omega = -  [\mathcal{L}_0^{\text{off}}]^{-1}\mathcal{J}_{1} +\left\{ -\left[\mathcal{L}_0^\text{off}\right]^{-1} + \left(\left[\mathcal{L}_0^\text{off}\right]^{-1} -\left[\mathcal{L}_0^\text{on}\right]^{-1}\right) e^{\tau_0\mathcal{L}_0^\text{off}}  + \left(\left[\mathcal{L}_0^\text{on}\right]^{-1} -\left[\mathcal{L}_0^\text{off}\right]^{-1}\right) e^{\tau_1\mathcal{L}_0^\text{on}}  e^{\tau_0\mathcal{L}_0^\text{off}}    \right\}\mathcal{J}_{-1} ~.
\end{equation}
Finally, since we have the operator $\Omega$, the steady state of the system is its eigenvector with eigenvalue 1, and the population of the ground and excited state are the diagonal elements of $\bar{\rho}_{ss}$.

The population of the excited state is then given by $P_e = \frac{1}{1+\xi}$, and one finds that
\begin{equation}
\xi = \frac{(\bar{N} + 1)}{\bar{N} \, u \, \omega_r^2 \, \delta}
\Bigg(
u \, \omega_r^2 \, \delta 
- 4 (1 + \bar{N}) \lambda^2 \omega_r^2 e^{- \bar{N} \tau_0 \gamma} 
 - \psi\Bigg)~,
\end{equation}
where
\begin{equation}
     \psi  =   2 \lambda^2 e^{- (\bar{N} \tau_0 + \tfrac{1}{2} u \tau_1) \gamma}
\left[
4 \delta 
+ 8 u \lambda^2 \left(e^{\frac{\tau_1 \omega_r}{2} } + e^{- \frac{\tau_1 \omega_r}{2} } \right)
- (\bar{N}+1) \gamma u \left( 
\omega_r \left(e^{\frac{\tau_1 \omega_r}{2} } - e^{- \frac{\tau_1 \omega_r}{2} } \right)
+ \gamma \left(e^{\frac{\tau_1 \omega_r}{2} } + e^{- \frac{\tau_1 \omega_r}{2} } \right)
\right)
\right]~,
\end{equation}
and
\begin{eqnarray}
    u &=& 2\bar{N}+1~,\\
    \delta &=& \bar{N}(\bar{N}+1) \gamma^2+4\lambda^2~,\\
    \omega_r &=& \sqrt{\gamma^2-16\lambda^2}~.
\end{eqnarray}
Note that when $\gamma \geq 4\lambda$, then $\omega_r\in \mathbb{R}$, and the exponential terms become hyperbolic functions. On the other hand, when $\gamma<4\lambda$, then $\omega_r = i \sqrt{16\lambda^2-\gamma^2}$ is a pure imaginary number, and these terms become a sine and cosine.

Now, let us outline the derivation of the main results for the example presented in the main text. First, we compute the optimal time $\tau_1^{\text{opt}}$ during which the external drive should remain on to maximize $P_e$. Maximizing $P_e$ is equivalent to minimizing $\xi$. Note that the only term in $\xi$ that depends on $\tau_1$ is proportional to $\psi$. Therefore, by solving the equation $\partial_{\tau_1} \psi = 0$, we obtain the optimal duration $\tau_1^{\text{opt}}$ as a function of $\bar{N}$, $\gamma$, $\lambda$, and $\tau_0$. Once $\tau_1^{\text{opt}}$ is determined, we set $\tau_1 = \tau_1^{\text{opt}}$ in all subsequent steps. Surprisingly, we found that the optimal duration of the drive depends only on the ratio $\gamma/\lambda$, and remains the same regardless of the temperature or the feedback time delay.

We can also determine the population inversion points by solving the equation $\xi = 1$, which corresponds to the condition $P_e = P_g = 1/2$. The solution of this equation yields the critical value $\bar{N}_c$ (or, equivalently, the critical temperature) for a given ratio $\gamma/\lambda$, such that $P_e > P_g$ for any $\bar{N} < \bar{N}_c$. We show numerically that in the ideal case with zero delay ($\tau_0 = 0$), the equation $\xi = 1$ admits a solution with $\bar{N}_c \geq 0$ only when $\gamma/\lambda < 1.145$.

Finally, to analyze the effect of feedback time delay on $P_e$, we consider the low-temperature limit $\bar{N} \rightarrow 0$, effectively removing thermal effects. We then compute the maximum time delay $\tau_0^{\text{max}}$ by solving the equation 
$\lim_{\bar{N}\rightarrow0} \xi = 1$ for $\tau_0$, given $\gamma$, $\lambda$, and $\tau_1^{\text{opt}}$. This yields the maximum delay $\tau_0^{\text{max}}$ for which $P_e = P_g$. Consequently, for any $\tau_0 < \tau_0^{\text{max}}$, we have $P_e > P_g$ provided $\gamma/\lambda < 1.145$.

\subsection{C. Example 2: reverting quantum transitions}

\begin{figure}
    \centering
      \includegraphics[scale=0.32]{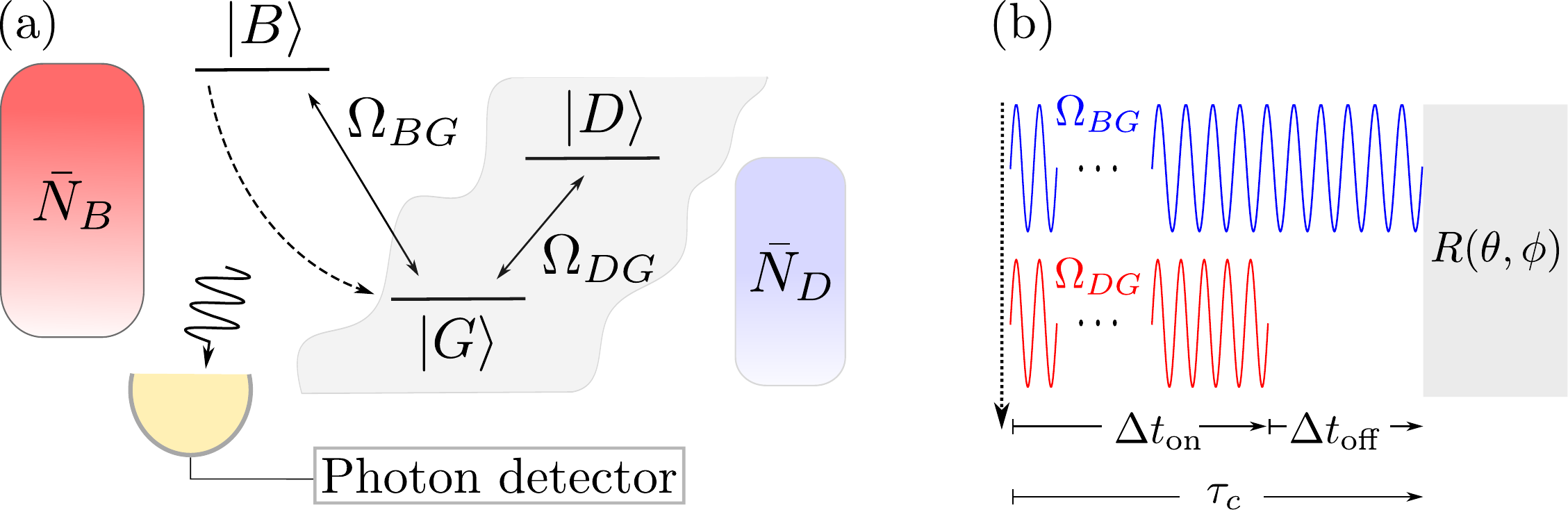}
     \caption{
   Feedback protocol used in Example~2.
    {(a)} Three-level system with a hidden transition~\cite{Minev2019CatchingReverseQuantumJump} between the ground $\ket{G}$ and dark $\ket{D}$ states (shaded region), minimally coupled to an effective thermal bath. The states $\ket{B}$ and $\ket{G}$ are coupled to a second thermal bath, and the incoherent jump $\ket{B}\to\ket{G}$ can be monitored via a photon detector.
    {(b)} The vertical line indicates the last detected jump. The system evolves under drives $\Omega_{BG}$ (blue) and $\Omega_{DG}$ (red). After a time $\tau_c$ has elapsed since the last jump, a pulse $R(\theta,\phi)$ is applied in the $\{\ket{G},\ket{D}\}$ subspace. The drive $\Omega_{DG}$ can either remain on during the full interval $\tau_c$ (as in Example~2) or be turned off after a time $\Delta t_{\text{on}}$.
    }
     \label{fig: example2}
\end{figure}

Let us consider the same system experimentally implemented in Ref.~\cite{Minev2019CatchingReverseQuantumJump}, corresponding to a superconducting artificial three-level atom (see Fig.~\ref{fig: example2}(a)).
The system consists of a ground state $\ket{G}$, a dark state $\ket{D}$, and a bright state $\ket{B}$.
The dark state $\ket{D}$ is engineered to minimally couple to any dissipative environment and measurement apparatus, with the states $\ket{D}$ and $\ket{G}$ coupled to an effective thermal bath of strength $\gamma_D$ and occupation $\bar{N}_D$, analogous to the qubit setup in Example~1.
A weak Rabi drive $\Omega_{DG}$ couples $\ket{G}$ and $\ket{D}$, while a stronger drive $\Omega_{BG} \gg \Omega_{DG}$ couples $\ket{B}$ and $\ket{G}$.
Similarly, the states $\ket{B}$ and $\ket{G}$ are coupled to a second bath characterized by $\gamma_B$ and $\bar{N}_B$.
This second bath induces incoherent transitions $\ket{B}\!\to\!\ket{G}$ that can be tracked using a photon-counting scheme, experimentally realizable through the monitoring of intermittent fluorescence from the bright state $\ket{B}$ in trapped ions~\cite{PhysRevLett.57.1699,PhysRevLett.57.1696,PhysRevLett.56.2797,PhysRevLett.54.1023}.
In this scheme, each incoherent jump $\ket{B}\!\to\!\ket{G}$ corresponds to a ``click" in the photon detector.

We assume that the coupling between $\ket{B}$ and $\ket{G}$ and the thermal bath dominates over the Rabi drives, $\gamma_{B}(\bar{N}_B+1)\gg \Omega_{BG}\gg \Omega_{DG}$. 
In this regime, coherent excitation from $\ket{G}$ to $\ket{B}$ is rapidly followed by an incoherent decay (jump) $\ket{B}\!\to\!\ket{G}$ emitting a photon. Consequently, photon detection signals population in $\ket{G}$, whereas the absence of photons indicates coherent evolution within the $\{\ket{G},\ket{D}\}$ subspace.
This system is described by the following quantum master equation [Eq.~\eqref{QME}]
\begin{eqnarray}
\label{eq: master equation of the three level system}
    \partial_t\rho_t =\mathcal{L}\rho_t = -i[H_\text{drive},\rho_t] +&\gamma_B \bar{N}_B \mathcal{D}\left[\ket{B}\bra{G}\right]\rho_t +\gamma_B (\bar{N}_B+1) \mathcal{D}\left[\ket{G}\bra{B}\right]\rho_t \\
     +&~\gamma_D \bar{N}_D \mathcal{D}\left[\ket{D}\bra{G}\right]\rho_t +\gamma_D (\bar{N}_D+1) \mathcal{D}\left[\ket{G}\bra{D}\right]\rho_t\nonumber ~,
\end{eqnarray}
where  $\mathcal{D}[L] \rho \equiv L\rho L^\dagger - \frac{1}{2} \{L^\dagger L, \rho\}$ is the Lindblad dissipator. 
The jump operators $L_B^{-} = \sqrt{\gamma_B \bar{N}_B}\ket{B}\bra{G}$ and $L_B^{+} = \sqrt{\gamma_B (\bar{N}_B+1)}\ket{G}\bra{B}$ describe, respectively, photon absorption from and emission into the thermal bath. Analogously, $L_D^{-} = \sqrt{\gamma_D \bar{N}_D}\ket{D}\bra{G}$ and $L_D^{+} = \sqrt{\gamma_D (\bar{N}_D+1)}\ket{G}\bra{D}$ represent the corresponding processes for the $\{\ket{G},\ket{D}\}$ manifold.
The drive Hamiltonian $H_{\text{drive}}$ reads
\begin{equation}
    H_{\text{drive}} = i \frac{\Omega_{BG}}{2} (\ket{B}\bra{G} -\ket{G}\bra{B} ) + i \frac{\Omega_{DG}}{2} (\ket{D}\bra{G} -\ket{G}\bra{D} ) ~.
\end{equation}

In this example, we consider the detection of only the thermal emission corresponding to the jump $\ket{B}\!\to\!\ket{G}$, described by $L_B^{-}$. Accordingly, the measurement outcome $x_t$ from the quantum jump monitoring at time $t$ takes the value $x_t = -1$ for a detected jump $\ket{B}\!\to\!\ket{G}$ and $x_t = 0$ for no-jump evolution.
The jump channel is
\begin{equation}
    \mathcal{J}_{-} \rho \equiv L_B^{-} ~\rho~ (L_B^{-})^\dagger = \gamma_B(\bar{N}_B+1)\bra{B}\rho\ket{B}\ket{G}\bra{G}~,
\end{equation}
and the no-jump Liouvillian is given by
\begin{equation}
    \mathcal{L}_0 \equiv \mathcal{L} - \mathcal{J}_{-}~,
\end{equation}
where $\mathcal{L}$ is defined by Eq.~\eqref{eq: master equation of the three level system}.
Consequently, when a jump $\ket{B}\!\to\!\ket{G}$ is detected, the system evolves under the jump channel $\mathcal{J}_{-}$; otherwise -- when no photon is detected in the bath coupling $\ket{B}$ and $\ket{G}$ -- it evolves under the no-jump Liouvillian $\mathcal{L}_0$.
The jump channel $\mathcal{J}_{-}$ defines a \emph{renewal process}~\cite{Tutorial}: regardless of the initial state $\rho$, the output $\mathcal{J}_{-}\rho$ is proportional to the pure state $\ket{G}\bra{G}$. In other words, $\mathcal{J}_{-}$ always resets the system to the ground state $\ket{G}$, independent of its prior evolution. We will return to this point shortly.

The feedback protocol in Example~2 is as follows (see Fig.~\ref{fig: example2}(b)): if the incoherent transition $\ket{B}\to\ket{G}$ is detected (via photon detection) and no further jumps occur for a time $\tau_c$, a pulse is applied in the $\{\ket{G},\ket{D}\}$ subspace, corresponding to the unitary transformation $R(\theta,\phi)$ defined as
\begin{equation}
    R(\theta,\phi) \rho = e^{i \frac{\theta}{2} \hat{n}\cdot \vec{\sigma}}~\rho~ e^{-i \frac{\theta}{2} \hat{n}\cdot \vec{\sigma}}~,
\end{equation}
and effectively represents a rotation in the $\{\ket{G},\ket{D}\}$ subspace.
The unit vector $\hat{n} = (\cos\phi, \sin\phi, 0)$ defines the rotation axis on the $DG$-Bloch sphere, and $\theta$ is the rotation angle. For example, $\theta = \pi$ and $\phi = \pi/2$ corresponds to a rotation of $\pi$ around the $y$-axis of the $DG$-Bloch sphere. 
For simplicity, in Example~2 we consider both drives $\Omega_{DG}$ and $\Omega_{BG}$ active during the interval $\tau_c$, but the case where $\Omega_{DG}$ is turned off after a time $\Delta t_{\text{on}}$ (Fig.~\ref{fig: example2}(b)) can also be treated using our formalism.

This feedback protocol is a time-dependent, jump-based scheme that uses the last detected jump and the elapsed time since its occurrence. Consequently, it is fully described by Result~2.
In particular, because the feedback action consists of an instantaneous rotation $R(\theta,\phi)$, the jump operators remain unchanged, and the system's state evolves according to
\begin{equation}
\label{eq: unconditional state evolution of Ex2}
    \bar{\rho}_t = G(\bar{k},t) \bar{\rho}_0+ \sum_{k\in\Sigma}\int_0^t d\tau G(k,\tau)\mathcal{J}_k\bar{\rho}_{t-\tau}~,
\end{equation}
where $\bar{\rho}_t$ is the state of the system at time $t$, $\Sigma$ is the set of monitored jumps, $G(k,\tau)$ is the no-jump propagator, and $\mathcal{J}_k$ are the jump channels.
We consider a generic initial condition in which the system is prepared in the state $\bar{\rho}_0$ and the jump memory is initialized in the state $\bar{k}$.
In this case, we detect only a single jump channel, $\mathcal{J}_{-}$, associated with the transition $\ket{B}\to\ket{G}$. Hence, the set $\Sigma$ contains only one symbol (denoted $-1$), and we can omit the jump index, keeping only the dependence on the elapsed time $\tau$. The system state then evolves according to $\bar{\rho}_t =G(t) \bar{\rho}_0+ \int_0^t d\tau \, G(\tau)\, \mathcal{J}_{-} \bar{\rho}_{t-\tau}$.

The no-jump propagator $G(\tau)$ describes the evolution in the absence of jumps, including the application of the pulse at time $\tau_c$. It is therefore given by
\begin{equation}
\label{eq: no jump propagator Ex2}
    G(\tau) = \begin{cases}
        e^{\tau\mathcal{L}_0}~,~\text{if }\tau<\tau_c~,\\
        e^{(\tau-\tau_c)\mathcal{L}_0}R(\theta,\phi)e^{\tau_c\mathcal{L}_0}~,~\text{if }\tau\ge \tau_c~.
    \end{cases}
\end{equation}
In other words, $G(\tau)$ describes the evolution of the system conditioned on no jumps $\ket{B}\to\ket{G}$. For $\tau < \tau_c$, the system evolves according to the no-jump Liouvillian $\mathcal{L}_0$. At $\tau = \tau_c$, the pulse is applied, and for $\tau > \tau_c$, the system continues its no-jump evolution.
We can verify that the no-jump Liouvillian $\mathcal{L}_0$ has eigenvalues with negative real parts. Consequently, $G(\tau)$ decays to zero as $\tau \to \infty$. Taking the limit $t\to\infty$ in Eq.~\eqref{eq: unconditional state evolution of Ex2}, we can extend the integral's upper limit to $+\infty$ and replace $\bar{\rho}_{t-\tau} \to \bar{\rho}_{ss}$ inside the integral, similarly to the discussion in Sec.~\textcolor{blue}{S.III}. This yields the steady-state equation $\bar{\rho}_{ss} = \Omega \bar{\rho}_{ss}$, where, in this example, $\Omega = \int_0^\infty d\tau\, G(\tau)\mathcal{J}_{-}$.
Using the explicit form of $G(\tau)$, the integral can then be evaluated, and we have 
\begin{equation}
\label{eq: steady state super operator of ex2}
    \Omega = \left[\mathcal{L}_0\right]^{-1}\left(e^{\tau_c \mathcal{L}_0} - \mathbb{1} - R(\theta,\phi)e^{\tau_c\mathcal{L}_0}\right)\mathcal{J}_{-},
\end{equation}
where $\left[\mathcal{L}_0\right]^{-1}$ is the inverse of the no-jump Liouvillian, and $\mathbb{1}$ is the identity super-operator. 
By employing the vectorization technique~\cite{Tutorial}, the super-operators are represented as matrices, and the states as vectors. In this representation, the steady state $\bar{\rho}_{ss}$ corresponds to the eigenvector of $\Omega$ with eigenvalue $1$, satisfying $\Omega \bar{\rho}_{ss} = \bar{\rho}_{ss}$.

Once the feedback steady state $\bar{\rho}_{ss}$ is obtained, it directly provides the jump statistics in the long-time regime for Example~2. According to Result~2, for a feedback that does not modify the jump operators, the memory-resolved state is given by $ \varrho_t(k,\tau) = 2 \delta(t-\tau)\delta_{k,\bar{k}}G(k,\tau)\bar{\rho}_0+G(k,\tau)\mathcal{J}_k\bar{\rho}_{t-\tau}$.
In this case, the memory-resolved steady state, denoted by $\varrho_{ss}(\tau)$, reads
\begin{equation}
\label{eq: memory-resolved steady state of Ex2}
    \varrho_{ss}(\tau) = G(\tau)\mathcal{J}_{-}\bar{\rho}_{ss} = \gamma_{B}(\bar{N}_B+1)\bra{B}\bar{\rho}_{ss}\ket{B}G(\tau)\left(\ket{G}{\bra{G}}\right)~,
\end{equation}
where $\bra{B}\bar{\rho}_{ss}\ket{B}\equiv P_B$ is the population of $\ket{B}$ in the steady state. 
Recall that $\varrho_{ss}(\tau)$ provides the probability density function of the counting memory $\tau_t$ in the stationary regime. In other words, $\mathrm{Tr}[\varrho_{ss}(\tau)] \equiv P_{ss}(\tau)$ is the stationary probability distribution that a time $\tau$ has elapsed since the last detected jump. This directly connects the feedback dynamics to measurable quantities.
For instance, the moments of the random time $\Delta t$ between jumps $\ket{B}\to \ket{G}$ are given by~\cite{Tutorial} $\braket{(\Delta t)^n} = (-1)^n n!
~\mathrm{Tr}\left( [\mathcal{L}_0]^{-n} \bar{\rho}_{ss} \right)$.
Note that $P_{ss}(\tau)$ is properly normalized:
\begin{equation}
    \int_0^\infty d\tau \, P_{ss}(\tau) = 
    \mathrm{Tr}\left[ \int_0^\infty d\tau \, \varrho_{ss}(\tau) \right] = 
    \mathrm{Tr}\left[ \bar{\rho}_{ss} \right] = 1~,
\end{equation}
where we used that $\int_0^\infty d\tau \, \varrho_{ss}(\tau) = \bar{\rho}_{ss}$.

The memory-resolved state provides the state of the system conditioned to a given memory realization. For this particular example, note that we can write
\begin{equation}
\label{eq: ensemble and memory states}
    \bar{\rho}_{ss} = \int_0^\infty d\tau~ \varrho_{ss}(\tau) = \int_0^{\infty}d\tau~ P_{ss}(\tau) \rho(\tau)~,
\end{equation}
where we define
\begin{equation}
\label{eq: def of memory states}
    \rho(\tau) \equiv \frac{\varrho_{ss}(\tau)}{\mathrm{Tr}[\varrho_{ss}(\tau)]}~.
\end{equation}
Hence, Eq.~\eqref{eq: ensemble and memory states} shows that the steady state $\bar{\rho}_{ss}$ can be interpreted as an ensemble $\{P_{ss}(\tau), \rho(\tau)\}$, with the states $\rho(\tau)$ defined in Eq.~\eqref{eq: def of memory states}. 
These states, which we refer to as \emph{memory-conditioned} states, represent the system's state conditioned on a specific realization of the memory function. 
From Eq.~\eqref{eq: memory-resolved steady state of Ex2}, the memory-conditioned state reads
\begin{equation}
\label{eq: memory-conditioned state of Ex2}
    \rho(\tau) = \frac{G(\tau)(\ket{G}\bra{G})}{\mathrm{Tr}\left[G(\tau)(\ket{G}\bra{G})\right]}~.
\end{equation}
Therefore, since the jump channel $\mathcal{J}_{-}$ defines a renewal process, the memory-conditioned state depends only on the ground state $\ket{G}$, to which the system is reset after the jump, and on the time interval during which no-jump evolutions are observed.

The experimental realization in Ref.~\cite{Minev2019CatchingReverseQuantumJump} implemented the same feedback protocol described in Example~2 using the three-level system shown in Fig.~\ref{fig: example2}(a). In their experiment, the jump $\ket{B}\to\ket{G}$ was not detected directly; instead, the bright state $\ket{B}$ was coupled to a cavity, and its population was inferred via homodyne detection. By monitoring the photon number in the cavity, one can determine whether the system is in $\ket{B}$ or in a superposition of $\ket{G}$ and $\ket{D}$. 
Nevertheless, a photon-counting detection scheme that tracks the incoherent jump $\ket{B}\to\ket{G}$ (as considered in Example~2~\cite{PhysRevLett.57.1699,PhysRevLett.57.1696,PhysRevLett.56.2797,PhysRevLett.54.1023}) provides a first-order approximation of these experimental results, as discussed in the supplementary material of Ref.~\cite{Minev2019CatchingReverseQuantumJump}.
In their experiment, they reported the populations of the system conditioned on no jumps $\ket{B}\to\ket{G}$ occurring during a time $\tau_c$, which corresponds to the instant of pulse application. In terms of our formalism, this is precisely the population of the memory-conditioned state $\rho(\tau_c)$. Assuming that the drive $\Omega_{DG}$ remains on throughout the interval $\tau_c$, this state is given by
\begin{equation}
\label{eq: nature-conditional-state1 - SupMat}
\rho(\tau_c) = \frac{R(\theta,\phi)~ e^{\tau_c \mathcal{L}_0}\left(\ket{G}\bra{G}\right)}{\mathrm{Tr}\left[R(\theta,\phi) ~e^{\tau_c \mathcal{L}_0}\left(\ket{G}\bra{G}\right)\right]},
\end{equation}
where we used Eq.~\eqref{eq: no jump propagator Ex2} of $G(\tau_c)$. We show the populations of $\rho(\tau_c)$ in the End Matter of the main text, as a function of both the waiting time $\tau_c$ and the pulse direction angle $\phi$.

The theoretical description in \cite{Minev2019CatchingReverseQuantumJump} relied on stochastic simulations of trajectories based on samples of the homodyne signal. Accessing the steady-state regime typically requires simulating a large number of long-time trajectories to gather meaningful statistics, which can be computationally expensive and does not yield analytical expressions, thereby limiting direct physical insight. 
In contrast, our framework provides an analytical approach to the same feedback scenario, yielding the feedback steady state directly from a simple eigenvector equation. Moreover, it gives systematic access to the jump statistics in the stationary regime under feedback, which are difficult to extract from trajectory-based simulations and were previously inaccessible with other theoretical descriptions of jump-based feedback.
Overall, our method offers a direct and comprehensive description of the feedback dynamics, providing analytical expressions that connect naturally to experimentally measurable quantities.

\section{S.V. Deriving previous results as particular cases of the Result 1}
\label{app:00000000E}
In this section, we show how the previous results \cite{PatrickQFPME, WisemanMilburn1993_Homodyne,Wiseman1994_Feedback,FPTLandi} can be seen as particular cases of the Result 1, given by the deterministic equation 
\begin{equation}
    \label{eq: deterministic eq 1 Appendix}
    \varrho_{n+1}(y) = \sum_{x',y'} \delta_{y, f_{n+1}(x',y')} \mathcal{M}_{x'}(y') \varrho_n(y')~,
\end{equation}
for a given stochastic memory $y_n$ that satisfies the causality equation $y_{n+1} = f_{n+1}(x_{n+1},y_n)$ for some function $f_n(x,y)$, where $x_n$ is the outcome of the detection at time $t_n = n \delta t$. 
In each case, the feedback-measurement protocol is fixed by defining the instruments $\mathcal{M}_x$ and the adequate memory function, as described in Table \ref{table: summary of results}. 
These previous works were developed using different approaches and were limited to specific choices of measurement schemes and memory functions. In contrast, our results provide a unified and general framework that describes arbitrary feedback dynamics, free from such restrictions and paving the way for more sophisticated feedback schemes.

\subsection{A. Low-pass memory and weak Gaussian 
measurements: Fokker-Planck master equation }
Let us consider a sequential measurement described by the instruments $\mathcal{M}_x$, with a low-pass memory $y_n$ defined as
\begin{equation}
\label{eq: low pass filter}
y_n = \sum_{j=1}^n \alpha p^{n-j} x_j~,
\end{equation}
where $\alpha$ and $p$ are arbitrary parameters, and $x_n$ is the outcome of the measurement at time $t_n = n \delta t$. From this definition, we can see that
\begin{equation}
    y_n = \alpha x_n+ p y_{n-1},
\end{equation}
then the low-pass memory satisfies the causality condition $y_n = f(x_n,y_{n-1})$, where $f(x,y) = \alpha x +p y$. From the Result 1, the corresponding deterministic equation of the memory-resolved state is given by
\begin{eqnarray}
    \varrho_{n+1}(y) = \sum_{x',y'} \delta_{y ,\alpha x' +p y'} \mathcal{M}_{x'}(y') \varrho_n(y')~.
\end{eqnarray}
For $p\neq0$, we can solve the sum over $y'$ by using the Kronecker delta, and the deterministic equation becomes
\begin{equation}
    \label{eq: deterministic equation for the low-pass filter}
    \varrho_n(y) = \sum_{x} \mathcal{M}_{x} ((y-\alpha x)/p)\varrho_{n-1}\big((y-\alpha x)/p\big).
\end{equation}
It is worth to mention that the case $p = 0$, where the memory corresponds to the actual detection, will be considered soon. Therefore, Eq.~\eqref{eq: deterministic equation for the low-pass filter} describes a general feedback-measurement protocol where the stochastic memory is given by Eq.~\eqref{eq: low pass filter}.

From now on, let us consider a sequential measurement described by the set of Kraus operators $\{K_x\}_{x \in \mathbb{R}}$, where
\begin{equation}
\label{eq: kraus operators of gaussian measurements}
    K_x = \left(\frac{2\lambda \delta t}{\pi}\right)^{\frac{1}{4}}e^{-\lambda \delta t (x-A)^2}~,
\end{equation}
$A$ is a given observable of the system, and $\lambda$ is the measurement strength~\cite{weakMeasurements1}. These operators $K_x$ describe a \textit{weak Gaussian measurement} of $A$, and the corresponding instrument is defined as $\widetilde{\mathcal{M}}_z\rho \equiv K_z\rho K_z^\dagger$. Now, let us consider that the system evolves dynamically according to a Liouvillian $\mathcal{L}$ between two detections. Hence, the total instrument $\mathcal{M}_z$ that describes this scenario is defined as
\begin{equation}
    \mathcal{M}_z(y_n) \equiv \widetilde{\mathcal{M}}_z e^{\delta t\mathcal{L}(y_n)}~,
\end{equation}
where the feedback encoded in the stochastic memory $y_n$ can affect the dynamic of the system through $\mathcal{L}(y_n)$. Therefore, from Eq.~\eqref{eq: deterministic equation for the low-pass filter}, we have 
\begin{equation}
\label{eq: deterministic equation for low-pass filter + Gaussian measurements}
    \varrho_n(y) = \int_{-\infty}^{+\infty} dx~   \widetilde{\mathcal{M}}_x e^{\mathcal{L}\left( \frac{y-\alpha x}{p} \right)\delta t} \varrho_{n-1}\left( \frac{y-\alpha x}{p} \right)~,
\end{equation}
where the sum is transformed into an integral because the outcomes $x$ are real numbers. 

In the next step we take the limit $\delta t \rightarrow0$ in the Eq.~\eqref{eq: deterministic equation for low-pass filter + Gaussian measurements}. Let us consider the change of variables defined as $\xi \equiv (y-\alpha x)/p$, and then $x = (y-p\xi)/ \alpha$. Also, we define $p = \exp{(-\gamma \delta t)} $ and $\alpha = \gamma \delta t$. Then $dx = - d\xi/ (\gamma \delta t)$. Therefore, the deterministic evolution becomes
\begin{equation}
    \varrho_n(y) = \int_{-\infty}^{+\infty} d\xi~\frac{1}{\gamma \delta t}~   \widetilde{\mathcal{M}}_{\left( \frac{y-e^{-\gamma \delta t}\xi}{\gamma \delta t} \right)} e^{\mathcal{L}\left(\xi \right)\delta t} \varrho_{n-1}\left(\xi \right)~.
\end{equation}
Now, we can define
\begin{equation}
    \Omega(y|\xi) \equiv\frac{1}{\gamma \delta t}~   \widetilde{\mathcal{M}}_{\left( \frac{y-e^{-\gamma \delta t}\xi}{\gamma \delta t} \right)}~,
\end{equation}
and then
\begin{equation}
\label{eq: low pass propagator equation}
    \varrho_n(y) = \int_{-\infty}^{+\infty} d\xi~\Omega(y|\xi)~ e^{\mathcal{L}\left(\xi \right)\delta t} \varrho_{n-1}\left(\xi \right)~.
\end{equation}

As shown in \cite{PatrickQFPME} (see supplemental material, Eq.~(S7)), we can expand the propagator $\Omega(y|\xi)~ e^{\mathcal{L}\left(\xi \right)\delta t}$ in first order of $\delta t$, resulting in the following expression
\begin{equation}
    \Omega(y|\xi) \rho = \delta(y-\xi)\rho + \delta t\left[ \lambda \delta(y-\xi) \mathcal{D}[A]\rho - \gamma \mathcal{A}(\xi) \rho \delta'(y-\xi) + \frac{\gamma^2}{8\lambda}\delta''(y-\xi) \rho  \right]\\ +~ \mathcal{O}(\delta t^2)~,
\end{equation}
where $\mathcal{D}[A]\rho = A\rho A^\dagger - 1/2 \{A^\dagger A,\rho\}$ is the Lindblad dissipator, and $\mathcal{A}(\xi)\rho:= \frac{1}{2}\{A-\xi,\rho\}$. Furthermore, $\delta'(x)$ and $\delta''(x)$ denote the first and second derivatives of the Dirac distribution $\delta(x)$, respectively. Therefore, applying this first order expansion of the propagator $\Omega(y|\xi)$ in Eq.~\eqref{eq: low pass propagator equation} and taking the limit $\delta t \rightarrow 0$, we finally have
\begin{equation}
\label{eq: FQME}
    \partial_t \varrho_t(y) = \mathcal{L}(y) \varrho_t(y) + \lambda \mathcal{D}[A]\varrho_t(y) - \gamma \partial_\xi \{\mathcal{A}(\xi) \varrho_t(\xi)\}|_{\xi = y} + \frac{\gamma^2}{8\lambda }\partial_\xi^2\varrho_t(\xi) |_{\xi = y}~,
\end{equation}
corresponding to the Fokker-Plank master equation of the memory-resolved state \cite{PatrickQFPME}. Note that, in the limit $\delta t \rightarrow0$, Eq.~\eqref{eq: low pass filter} becomes
\begin{equation}
    y_t = \int_0^t ds \gamma e^{-\gamma(t-s) } x_t~, 
\end{equation}
where $x_t$ is the outcome of the weak Gaussian measurement at time $t$. Therefore, we showed that Eq.~\eqref{eq: FQME} can be derived from Result 1 considering a sequential weak Gaussian measurement and a feedback-measurement protocol based on the low-pass memory. 

\subsection{B. Feedback conditioned on the present measurement result}

In the context of sequential detection, we consider feedback that depends solely on the most recent measurement outcome. 
This corresponds to a stochastic memory given by the current detection, $y_n = x_n$, and can be seen as a particular case of the low-pass memory with $p = 0$. 
In this protocol, the feedback does not rely on the prior history of the experiment, and we have $y_n = f(x_n,y_{n-1})$ with $f(x,y) = x$.
According to Result 1, the deterministic equation governing the memory-resolved state is then given by
\begin{equation}
    \varrho_{n+1}(y) = \sum_{x',y'} \delta_{y,x'} \mathcal{M}_{x'}(y') \varrho_n(y')~,
\end{equation}
and by performing the sum over $x'$, we obtain
\begin{equation}
    \label{eq: deterministic equation for the current resolved state}
    \varrho_{n+1}(y) = \sum_{y'}  \mathcal{M}_{y}(y') \varrho_n(y')~,
\end{equation}
where $y$ denotes a possible detection outcome at time $t_{n+1}$, and the sum over $y'$ accounts for all possible outcomes at time $t_n$. Equation~\eqref{eq: deterministic equation for the current resolved state} therefore describes a general feedback-measurement protocol based solely on the most recent detection. In what follows, we explore how this equation applies to specific measurement schemes, thereby recovering earlier results.

\subsubsection{1. Quantum Jumps}
Let us consider the application of a quantum channel to the system's state conditioned on quantum jump detection. The conditional stroboscopic evolution in this feedback-measurement scenario is described by
\begin{equation}
    \rho_n = \mathcal{F}(x)\left[\frac{V_x\rho_{n-1}V_x^\dagger}{\mathrm{Tr}[V_x\rho_{n-1}V_x^\dagger]}\right]~,
\end{equation}
where the Kraus operators $\{V_x\}_{x}$ describe the quantum jump detection process (see section \textcolor{blue}{S.I~C} for a quick review): $V_k = \sqrt{\delta t}L_k$ for $k \in \Sigma$, $V_0 = \mathbb{1} - i \delta t H_{\text{eff}}$, $H_{\text{eff}} = H - \frac{i}{2}\sum_{k\in \Sigma} L_k^\dagger L_k$ is the effective Hamiltonian, and $\Sigma$ is the set of all possible jumps. 
In this case, the quantum channel $\mathcal{F}(x)$, referred to as the \textit{feedback super-operator}, defines the feedback protocol based on the most recent quantum jump~\cite{Wiseman1994_Feedback}. 
The corresponding instrument for this sequential feedback-measurement protocol is given by $\mathcal{M}_x \rho = \mathcal{F}(x)[V_x(\rho)V_x^\dagger]$, where $x = 0$ represents a no-jump detection, and $x \in \Sigma$ corresponds to a quantum jump. 

Then, using Eq.~\eqref{eq: deterministic equation for the current resolved state}, we obtain the deterministic equation for the memory-resolved state:
\begin{equation}
    \varrho_{n}(k)=\sum_{x}\mathcal{F}(k)V_k\varrho_{n-1}(x)V_k^\dagger = \mathcal{F}(k)V_k\left[\sum_{x}\varrho_{n-1}(x)\right]V_k^\dagger = \mathcal{F}(k)V_k \bar{\rho}_nV_k^\dagger~,
\end{equation}
where we have used the identity $\bar{\rho}_n = \sum_x \varrho_n(x)$. Summing this equation over $k$, we finally obtain
\begin{equation}
\label{ew: deterministic equation for rho and feedback superoperator}
\bar{\rho}_n =  \sum_{k} \mathcal{F}(k)V_k \bar{\rho}_{n-1}V_k^\dagger. 
\end{equation}
Now, let us consider the following feedback super-operator
\begin{equation}
    \mathcal{F}(k) =
\begin{cases}
\mathbb{1} & \text{if } k = 0 \text{ (no-jump)}, \\
e^{\mathcal{K}(k)} & \text{if } k \in \Sigma \text{ (jump $k$)},
\end{cases}
\end{equation}
where $\mathcal{K}$ is an arbitrary Liouvillian super-operator. Hence, we can write the evolution of the uncoditional state as
\begin{equation}
\bar{\rho}_n = V_0\bar{\rho}_{n-1}V_0^\dagger +  \sum_{k \in \Sigma} \mathcal{F}(k)V_k \bar{\rho}_{n-1}V_k^\dagger.
\end{equation}
Finally, using the definition of the Kraus operators $V_k$, and taking the limit $\delta t \rightarrow 0$, we obtain 
\begin{eqnarray}
\label{eq: single jump feedback equation}
    \partial_t \bar{\rho}_t = -i [H,\bar{\rho}_t] + \sum_{k \in \Sigma} \left( e^{\mathcal{K}(k)} \left[L_k \bar{\rho}_tL_k^\dagger\right] - \frac{1}{2}\{L_k^\dagger L_k,\bar{\rho}_t\} \right)~,
\end{eqnarray}
corresponding to the master equation for feedback based on the present jump detection, as developed in \cite{Wiseman1994_Feedback}.

\subsubsection{2. Diffusion}

Let us consider another scenario where the feedback is based solely on the most recent measurement outcome, but now with diffusion. Considering a weak continuous Gaussian measurement of an hermitian observable $A$ described by the Kraus operators $K_z$ [Eq.~\eqref{eq: kraus operators of gaussian measurements}], the outcome of the measurements follows a stochastic equation given by
\begin{equation}
    z(t) = \frac{1}{2 \sqrt{\lambda}\delta t}dW(t) + \left< A \right>_c~,
\end{equation}
where $ \left< A \right>_c = \mathrm{Tr}[A \rho_c(t)]$, $dW$ is a Wiener increment, and $\rho_c(t)$ is the conditional state at time $t$. The stochastic master equation for $\rho_c$ is given by \cite{weakMeasurements1}
\begin{equation}
\label{eq: Belevkin equation}
    d\rho_c(t)  = \delta t \mathcal{L}\rho_c(t) + \lambda \delta t \mathcal{D}[A]\rho_c(t) + \sqrt{\lambda}dW\{A-\left< A \right>_c,\rho_c(t)\}.
\end{equation}
In this case, the feedback protocol is defined as follows: the system evolves dynamically from $t_{n-1}=(n-1)\delta t$ to $t_n = n \delta t$, and we apply a weak Gaussian measurement of $A$ at this point, resulting in the outcome $z_n$. Hence, the feedback is based on the most recent outcome of the detection through the feedback super-operator $\mathcal{F}(z)$. For simplicity, we consider a unitary evolution between two measurements. 

The total instrument that describes this discrete evolution is given by $\mathcal{F}(z)\widetilde{\mathcal{M}}_z e^{\delta t \mathcal{L}}$, where $\mathcal{L}\rho = -i[H,\rho]$ represents the unitary evolution between $t_{n-1}$ and $t_n$, $\widetilde{\mathcal{M}}_z \rho \equiv K_z \rho K_z^\dagger$ represents the weak Gaussian measurement, and $\mathcal{F}(z)$ is the feedback super-operator based on the measurement outcome. Following a similar approach as we developed for quantum jumps in Eq.~\eqref{ew: deterministic equation for rho and feedback superoperator}, the deterministic equation for the unconditional state in this case is given by
\begin{equation}
\bar{\rho}_n =  \int_{-\infty}^{\infty}dz \mathcal{F}(z)\widetilde{\mathcal{M}}_ze^{\delta t\mathcal{L}}\bar{\rho}_{n-1}, 
\end{equation}
where the sum was replaced by a integral since the outcomes of the Gaussian measurements are continuous real numbers.

Now, let us consider a particular feedback super-operator. Following the reference~\cite{WisemanBook}, let us define 
\begin{equation}
    \mathcal{F}(z) = e^{ 2\sqrt{\lambda} z \delta t \mathcal{K} }~,
\end{equation}
where $\mathcal{K}\rho \equiv -i[F,\rho]$ is a super-operator that represents a reversible transformation for a given hermitian operator $F$. Hence, the deterministic equation becomes 
\begin{equation}
    \bar{\rho}_n = \int_{-\infty}^{+\infty} dz e^{2\sqrt{\lambda}z\delta t\mathcal{K}}\widetilde{\mathcal{M}}_z e^{\delta t \mathcal{L}}\bar{\rho}_{n-1}~.
\end{equation}
The next step corresponds to expand the deterministic equation on first order of $\delta t$. For the feedback super-operator, we have \cite{WisemanBook}
\begin{equation}
    \mathcal{F}(z) = 1 + 2 \sqrt{\lambda} z \mathcal{K} \delta t + \frac{\mathcal{K}^2}{2}\delta t~,
\end{equation}
and the unitary evolutions yields 
\begin{equation}
    e^{\delta t \mathcal{L}}\rho \approx (\mathbb{1} + \delta t \mathcal{L})\rho = \rho-i\delta t[H, \rho]~.
\end{equation}
Also, we can show that 
\begin{equation}
    \widetilde{\mathcal{M}}_z\rho = K_z\rho K_z^\dagger = \int_{-\infty}^{+\infty}\frac{d\omega}{2\pi} e^{\frac{-\omega^2}{8 \lambda \delta t}}e^{-i\omega z} \left[ e^{\frac{i \omega A}{2}} \rho e^{\frac{i \omega A}{2}} - \frac{\lambda \delta t}{2}\left(A^2 e^{\frac{i \omega A}{2}} \rho e^{\frac{i \omega A}{2}} + e^{\frac{i \omega A}{2}} \rho A^2 e^{\frac{i \omega A}{2}} - 2 A e^{\frac{i \omega A}{2}} \rho A e^{\frac{i \omega A}{2}}\right) \right]~,
\end{equation}
and keeping only terms of order $\delta t$ in the instrument $e^{2\sqrt{\lambda}z\delta t\mathcal{K}}\widetilde{\mathcal{M}}_z e^{\delta t \mathcal{L}}$, the deterministic equation becomes
\begin{equation}
    \bar{\rho}_n = \bar{\rho}_{n-1} + \lambda \delta t \mathcal{D}[A] \bar{\rho}_{n-1} -i \delta t [H,\bar{\rho}_{n-1}]  + \delta t \mathcal{D}[F]\bar{\rho}_{n-1}  - \sqrt{\lambda}\delta t i [F, A\bar{\rho}_{n-1} + \bar{\rho}_{n-1} A]~.
\end{equation}
Finally, taking the limit $\delta t\rightarrow0$ and writing $A = A^\dagger$, one finds
\begin{equation}
\label{eq: FB + diffusion}
   \partial_t \bar{\rho}_t =  -i[H,\bar{\rho}_t]+  \mathcal{D}\left[\sqrt{\lambda }A \right] \bar{\rho}_{t} +   \mathcal{D}[F]\bar{\rho}_{t}  -  i [F, \sqrt{\lambda}A\bar{\rho}_{t} + \bar{\rho}_{t} \sqrt{\lambda}A^\dagger]~.
\end{equation}

Eq.~\eqref{eq: FB + diffusion} coincides with the feedback master equation found in~\cite{WisemanMilburn1993_Homodyne} for ideal homodyne detection. In that case, one must make the identification $\sqrt{\lambda} A \rightarrow L$, where $L$ is a (generally non-Hermitian) jump operator. In homodyne detection, the diffusion arises from continuous measurements of the quadrature operator $x = L + L^\dagger$. Therefore, our result recovers the feedback master equation of~\cite{WisemanMilburn1993_Homodyne}, corresponding to diffusion induced by weak Gaussian measurements of a Hermitian operator $A$.

\subsection{C. Charge-based feedback}
Let us consider a continuous monitoring of quantum jumps, as defined in the main text. In this case, $x_n$ represents the outcome of the quantum jump detection at time $t_n = n\delta t$, where $x_n = 0$ is a no-jump detection, and $x_n = k \in \Sigma$ is a jump in the channel $k$. Let us introduce the following random variable 
\begin{equation}
    dN_{k,n} \equiv\delta_{x_n, k}~,
\end{equation}
where $\delta_{a,b}$ is the Kronecker delta. It implies that $dN_{k,n} = 1$ if we have a jump in the channel $k$ at time $t_n$, and $dN_{k,n} = 0$ otherwise. Furthermore, the total stochastic number of jumps in the channel $k$ until the time $t_n$ is
\begin{equation}
    N_{k,n} \equiv \sum_{j=1}^n dN_{k,j} = \sum_{j=1}^n \delta_{x_j,k}~.
\end{equation}
Hence, we can define the stochastic charge as~\cite{Tutorial}
\begin{equation}
\label{eq: stochastic charge}
    N_n \equiv \sum_{k\in\Sigma}\nu_k N_{k,n} = \sum_{k\in\Sigma}\sum_{j=1}^n \nu_k \delta_{x_j,k}~,
\end{equation}
corresponding to a linear combination of the total number of jumps in each channel $k$. In the limit $\delta t\rightarrow0$, we have
\begin{equation}
    N(t) =  \sum_{k\in\Sigma}\nu_k\int_{0}^t dt' \delta(x(t') - k)~,
\end{equation}
where $\delta(x)$ denotes the Dirac delta. In this way, we have defined the stochastic charge in terms of the measurement outcomes $x_n$ of the continuous monitoring of the quantum jumps. Note that this stochastic memory is causal, and it satisfies
\begin{equation}
\label{eq: recursive relation of the charge}
    N_n = \sum_{k\in\Sigma}\nu_k \delta_{x_n,k} + N_{n-1}.
\end{equation}

Now, let us consider a feedback protocol where the detections are based on the continuous monitoring of the quantum jumps, and the stochastic memory is the stochastic charge \eqref{eq: stochastic charge}. In this case, the instruments are given by
\begin{align}
    \mathcal{M}_0(N)\rho &= (1+\mathcal{L}_0(N) \delta t) \rho, 
    \\[0.2cm]
    \mathcal{M}_k(N)\rho &= \delta t \mathcal{J}_k(N) \rho,
\end{align}
where $\mathcal{J}_k\rho = L_k^{}(y) \rho L_k^\dagger(y)$ and
$\mathcal{L}_0(y)  = \mathcal{L}(y) - \sum_{k\in\Sigma} \mathcal{J}_k(y)$.
By applying Eq.~\eqref{eq: result 1} and \eqref{eq: recursive relation of the charge}, one finds
\begin{eqnarray}
    \varrho_{n+1}(N) &=& \sum_{x'} \sum_{N'} \delta_{N,\sum_{k\in\Sigma}\delta_{x',k}+N'}\mathcal{M}_{x'}(N')\varrho_n(N')\\
    &=& \sum_{N'}\delta_{N,N'}\mathcal{M}_0(N')\varrho_n(N') + \sum_{x'\in\Sigma} \sum_{N'} \delta_{N,\nu_{x'}+N'}\mathcal{M}_{x'}(N')\varrho_n(N')\\
    &=& \mathcal{M}_0(N)\varrho_n(N) + \sum_{x'\in\Sigma}\mathcal{M}_{x'}(N-\nu_{x'})\varrho_n(N-\nu_{x'})~.
\end{eqnarray}
Finally, using the definition of $\mathcal{M}_0$ and $\mathcal{M}_k$, and taking the limit of $\delta t \rightarrow0$, we have
\begin{equation}
\label{eq: FPT feedback equation}
    \partial_t \varrho_t(N) = \mathcal{L}_0(N)\varrho_t(N) + \sum_{k\in\Sigma} \mathcal{J}_k(N-\nu_k)\varrho_t(N-\nu_k)~.
\end{equation}
Eq.~\eqref{eq: FPT feedback equation} was previously derived in \cite{FPTLandi} using a completely different approach, encompassing any feedback protocol based on charge detection. We show here that this result follows directly from Eq.~\eqref{eq: result 1}.

\end{document}